\documentclass[3p]{elsarticle}
\usepackage[utf8]{inputenc}
\usepackage{amsmath}
\usepackage{amsfonts}
\usepackage{hyperref}
\usepackage{verbatim}
\usepackage{graphicx}% Include figure files
\usepackage[linesnumbered, ruled,vlined]{algorithm2e}
\usepackage{commath}  % for \dif

\usepackage{hyperref}

\begin{document}

\begin{frontmatter}

\title{Crack-path selection in phase-field models for brittle fracture}

\author[imtek]{W. Beck Andrews}
\ead{william.beck.andrews@gmail.com}

\author[imtek,livmats]{Lars Pastewka}

\address[imtek]{Department of Microsystems Engineering, University of Freiburg, Georges-K\"ohler-Allee 103, 79110 Freiburg, Germany}
\address[livmats]{Cluster of Excellence livMatS, Freiburg Center for Interactive Materials and Bioinspired Technologies, University of Freiburg, 79110 Freiburg, Germany}

\begin{abstract}
This work presents a critical overview of the effects of different aspects of model formulation on crack path selection in quasi-static phase field fracture.
We consider different evolution methods, mechanics formulations, fracture dissipation energy formulations, and forms of the irreversibility condition.
The different model variants are implemented with common numerical methods based on staggered solution of the phase-field and mechanics sub-problems via FFT-based solvers.
These methods mix standard approaches with novel elements, such as the use of bound-constrained conjugate gradients for the phase field sub-problem and a heuristic method for near-equilibrium evolution.
We examine differences in crack paths between model variants in simple model systems and microstructures with randomly heterogeneous Young's modulus.
Our results indicate that near-equilibrium evolution methods are preferable for quasi-static fracture of heterogeneous microstructures compared to minimization and time-dependent methods.
In examining mechanics formulations, we find distinct effects of crack driving force and the model for contact implicit in phase field fracture.
Our results favor the use of a strain-spectral decomposition for the crack driving force but not the contact model.
Irreversibility condition and fracture dissipation energy formulation were also found to affect crack path selection, but systematic effects were difficult to deduce due to the overall sensitivity of crack selection within the heterogeneous microstructures.
Our findings support the use of the AT1 model over the AT2 model and irreversibility of the phase field within a crack set rather than the entire domain.
Sensitivity to these differences in formulation was reduced but not eliminated by reducing the crack width parameter $\ell$ relative to the size scale of the random microstructures.

\end{abstract}


\end{frontmatter}

\section{Introduction}
Phase field fracture \cite{bourdin_numerical_2000,ambati_review_2015,wu_chapter_2020} is a leading tool for investigating fracture in engineered \cite{roters_damask_2019,bui_review_2021}, geological \cite{wilson_phase-field_2016}, and biological materials \cite{shen_novel_2019}.
By considering cracks as localized changes in a phase field variable, phase field fracture requires no explicit tracking of the crack front, and can thus simulate arbitrarily complex crack geometries.
Phase field fracture is straightforward to generalize to different physical scenarios, with variants for dynamic \cite{karma_phase-field_2001,bourdin_time-discrete_2011} and quasi-static \cite{bourdin_numerical_2000} fracture and extensions that include plasticity \cite{roters_damask_2019,alessi_comparison_2018} and a variety of other multi-physics phenomena \cite{wilson_phase-field_2016,bilgen_phase-field_2021,svolos_thermal-conductivity_2020}.
The development of phase field fracture over the last 20 years has even lead to a variety of different models for even the relatively simple case of quasi-static brittle fracture (see reviews \cite{ambati_review_2015,wu_chapter_2020}).
For researchers interested in using phase field fracture, systematic comparisons of these models are valuable in determining what is physically appropriate for their system.
However, such comparisons \cite{ambati_review_2015,kuhn_degradation_2015,linse_convergence_2017,tanne_crack_2018, de_lorenzis_nucleation_2021, zhang_assessment_2022} have so far focused on homogeneous systems, despite the growing number of phase field fracture studies that are explicitly interested in the effects of material heterogeneity \cite{chakraborty_multi-scale_2016,hansen-dorr_phase-field_2020,mesgarnejad_crack_2020,wang_modeling_2021,lotfolahpour_effects_2021}.
The present study seeks to address this gap by focusing on how different phase field fracture formulations affect crack paths in a set of randomly generated, elastically heterogeneous two-dimensional (2-D) microstructures.

Our interest in crack paths is motivated by the problem of predicting the geometry of fracture surfaces.
Fracture surfaces are known to exhibit self-affine scaling \cite{mandelbrot_fractal_1984,maloy_experimental_1992}, and understanding this geometrical scaling has been a goal of modeling and simulation efforts for over 30 years (see e.g., Ref.\ \cite{ponson_statistical_2016} for a review).
Models for fracture surface roughness in brittle materials \cite{larralde_shape_1995,ramanathan_quasistatic_1997,katzav_fracture_2007,ponson_statistical_2016} consider the evolution of a sharp crack via propagation laws based on solutions to the stress field around the crack obtained from linear elastic fracture mechanics (LEFM) \cite{zehnder_fracture_2012}.
In such models, the crack propagates in the direction indicated by the principle of local symmetry, in which the direction in which the stress intensity factor for mode II (in-plane shear) $K_\mathrm{II}$ is zero \cite{goldstein_brittle_1974,hodgdon_derivation_1993}, when Griffith's criteria \cite{griffith_vi_1921} is met in this direction.
That is, when the elastic energy $G$ that would be released by extending the crack by a unit distance exceeds a critical value $G_c$.
Other criteria for the direction and onset of crack growth exist, but differences between them are minimal for isotropic materials \cite{cotterell_slightly_1980,hutchinson_mixed_1991}.
The stress distributions obtained from LEFM enable analytical predictions \cite{larralde_shape_1995,ramanathan_quasistatic_1997,ponson_statistical_2016} and efficient simulations \cite{katzav_fracture_2007,lebihain_effective_2020}, but only for systems where the elasticity problem is analytically tractable, for example when elastic properties are uniform or their effects can be abstracted into a noise term acting on the crack path.

Phase field fracture is more general than these sharp-crack evolution models in that it can simulate crack nucleation and branching in addition to propagation, and it is not limited to systems where LEFM can be applied.
The formulation of Ref.~\cite{bourdin_numerical_2000} from which most contemporary phase field fracture formulations originate was proposed as a regularization of the variational approach to fracture \cite{francfort_revisiting_1998}, in which the crack growth criterion $G>G_c$ is recovered via variational principles from an energy functional containing both the stored elastic energy (dependent on the phase field and strain field) and the energy dissipated during propagation of the crack (dependent only on the phase field).
This fracture dissipation energy contains a diffuse interface approximation for the crack measure that $\Gamma$-converges as a crack width parameter $\ell$ goes to zero \cite{ambrosio_approximation_1990,braides_approximation_1998}.
This approximation was originally proposed by Ambrosio and Tortorelli \cite{ambrosio_approximation_1990} for the Mumford-Shah image segmentation problem.
The limit $\ell\to 0$ was also investigated via matched asymptotic analysis by Hakim and Karma \cite{hakim_laws_2009}, who confirmed agreement with the principle of local symmetry for propagation through isotropic materials and considered anisotropic fracture toughness via simulations.
In addition to describing crack propagation, phase field fracture models have been shown to describe crack nucleation in a way that accurately matches experimental systems with stress concentrations and singularities \cite{tanne_crack_2018}.
Phase field fracture has also been interpreted as a form of continuum damage model \cite{pham_gradient_2011,de_borst_gradient_2016}, which provides a physical interpretation to evolution of the phase field away from a crack.

While most studies point to agreement between phase field fracture and classical theories, there are certain scenarios and formulations that are known to result in behavior that is non-physical in the context of brittle fracture.
One example is the possibility for interpenetration of crack faces and crack nucleation in compression in the initial model of Bourdin et al.\ \cite{bourdin_numerical_2000}.
Multiple formulations were subsequently proposed to restrict the driving force for fracture to tensile or shear conditions and to enforce some form of elastic contact between crack faces \cite{amor_regularized_2009,lancioni_variational_2009,freddi_regularized_2010,miehe_thermodynamically_2010,zhang_assessment_2022}  (see also Ref.\ \cite{wu_chapter_2020} for a review).
These include non-variational formulations \cite{ambati_review_2015}, in which the governing equations for the strain field and phase field do not correspond to the same energy functional.
A second example concerns the form of the fracture dissipation energy in the original model of Ref.\ \cite{bourdin_numerical_2000}, in which the phase field evolves even at low stresses leading to the lack of a purely elastic phase prior to fracture \cite{pham_gradient_2011}.
An alternative formulation with an elastic phase leads to an improved description of crack nucleation compared to experiments \cite{tanne_crack_2018}.
A related third example is the irreversibility of crack growth: constraining the entire evolution of the phase field to be irreversible, as opposed to a crack set \cite{bourdin_numerical_2000,gerasimov_penalization_2019}, can lead to poor $\Gamma$-convergence \cite{linse_convergence_2017}.

In this work, we consider formulations of the elastic energy, fracture dissipation energy, and irreversibility condition as three `dimensions' in which models for quasi-static fracture can vary.
As a fourth `dimension', we also consider the method for evolving the phase field.
We consider three types of evolution method: minimization \cite{bourdin_numerical_2000,bourdin_numerical_2007,bourdin_variational_2008}, time-dependent evolution \cite{miehe_thermodynamically_2010,kuhn_continuum_2010}, and near-equilibrium (e.g., path-following \cite{vignollet_phase-field_2014,may_new_2016,singh_fracture-controlled_2016}) methods.
To our knowledge, our study is the first comprehensive comparison between all three of these evolution methods for quasi-static phase field fracture.

The different formulations considered in this work are simulated within a common numerical and computational framework.
Our solvers weakly couple the phase field and elasticity sub-problems, a relatively common approach in phase field fracture (see e.g., Refs.\ \cite{bourdin_numerical_2000,miehe_phase_2010,singh_fracture-controlled_2016}).
The phase field and elasticity sub-problems are discretized using spectral methods \cite{saranen_periodic_2002,zeman_finite_2017} that make use of fast Fourier transforms (FFTs), although our approach differs in certain technical aspects from previous FFT-based implementations \cite{chen_fft_2019,ernesti_fast_2020,pankowski_fourier_2020}.
Notably, we apply a bound-constrained conjugate gradients algorithm \cite{vollebregt_bound-constrained_2014} to solve for the phase field while constraining it to evolve irreversibly.
Our example of a path-following method is also novel for phase field fracture, and its attributes compared to previous methods \cite{vignollet_phase-field_2014,may_new_2016,singh_fracture-controlled_2016} will be briefly discussed.
Overall, this work is focused on comparing model formulations with respect to crack path selection, and other questions about the relative suitability of our methods are left to future work.

\section{Background \label{sec:background}}
Phase field fracture models simulate damage and fracture via the evolution of the phase field $\phi$ within the entire $d$-dimensional domain.
There are multiple approaches to determining this evolution, but they all at some level involve solving partial differential equations for $\phi$ and the displacement or strain field.
The phase field represents both local degradation of the elastic properties of the material and the dissipation of energy due to disruption of bonds in the material via damage or formation of a crack.
During fracture, evolution of the phase field becomes localized around one or more $(d-1)$-dimensional cracks.
In order to avoid healing damage or cracks that developed at previous steps, evolution of $\phi$ must be constrained to be irreversible, either throughout the entire domain \cite{miehe_phase_2010,pham_gradient_2011} or within a crack set where $\phi$ has reached some critical value \cite{bourdin_numerical_2000}.
Phase field fracture models are formulated such that $\phi$ varies smoothly between its fully damaged state (e.g., $\phi=1$) at a crack center and its value in the bulk material (e.g., $\phi=0$).
This regularity around the crack provides phase field fracture with a degree of independence from its spatial discretization \cite{de_borst_gradient_2016,linse_convergence_2017}, provided that the discretization elements are sufficiently small relative to the length scale over which $\phi$ decays.
The capability to simultaneously nucleate and evolve multiple cracks independently of the spatial discretization makes phase field fracture a promising method for investigating crack path selection.

\subsection{Free energy functional}

The usual starting point for describing phase field fracture models is a free energy functional.
For quasi-static brittle fracture, this functional has two parts,
\begin{equation}
    F[\phi, \mathbf u] := F_e[\phi, \mathbf u] + F_f[\phi],
\end{equation}
where $\mathbf u$ is the displacement vector, $F_e$ is the stored elastic energy, and $F_f$ is the energy dissipated during fracture.
Here and in the following we use bold-faced symbols for vectors and square brackets to indicate functional dependence.
The elastic energy $F_e[\phi,\mathbf u]$ is simply the integral over the domain of the elastic energy density $\psi(\phi,\mathbf{\varepsilon})$ for a material point with phase field $\phi$ and strain $\mathbf{\varepsilon}=\nabla_s \mathbf u$, where $\nabla_s$ denotes the symmetrized gradient $\nabla_s \mathbf u=(\nabla \mathbf u +\nabla^T \mathbf u)/2$,
\begin{equation}
\label{eq:Fe}
    F_e[\phi,\mathbf u] := \int_\Omega \psi\left(\phi,\mathbf{\varepsilon}(\nabla \mathbf u)\right) \dif \mathbf x,
\end{equation}
where $\Omega$ denotes the $d$-dimensional simulation domain $\Omega \subset \mathbb R^d$ and $\mathbf x \in \Omega$.
In the simplest choice for $\psi(\phi,\varepsilon)$, the classical small-strain elastic energy density for an isotropic solid is multiplied by a quadratic degradation function $h(\phi)=(1-\phi)^2$ \cite{bourdin_numerical_2000,pham_gradient_2011},
\begin{equation}
\label{eq:isotropic_en}
    \psi(\phi,\mathbf{\varepsilon}) = \frac{1}{2}\left(\lambda \mathrm{tr}(\mathbf{\varepsilon})^2 + 2\mu \sum_i^d \sum_j^d \varepsilon_{ij}^2\right) h(\phi),
\end{equation}
where $\mathrm{tr}(\mathbf{\varepsilon})$ is the trace of $\mathbf{\varepsilon}$ and $\lambda$ and $\mu$ are the Lam\'e parameters: $\mu=E/(2+2\nu)$ and $\lambda = E\nu/(1-\nu-2\nu^2)$ in terms of the Young's modulus $E$ and Poisson's ratio $\nu$.
The degradation function $h(\phi)$ is equal to unity in the undamaged state, $h(0)=1$, and zero in the fully damaged state at the crack center, $h(1)=0$.
It also has a derivative of zero at the fully damaged state, $h'(1)=0$, which means that there is no driving force for further increases in $\phi$ beyond $\phi=1$.
The elastic energy density $\psi(\phi,\mathbf{\varepsilon})$ is non-convex in $\phi$ and $\mathbf{\varepsilon}$ when they are considered together, but convex in each when the other variable is held constant.
The model in Eq.~\eqref{eq:isotropic_en} is referred to as isotropic because it does not distinguish between tensile and compressive strain states \cite{bourdin_numerical_2000,miehe_thermodynamically_2010}.
Thus, a crack in this model would be stress-free even under compressive strains where a contact stress would be expected physically.

In order to account for the asymmetric response of a crack to tension vs.\ compression, the elastic energy density is typically split into two terms: $\psi^+_0(\mathbf{\varepsilon})$, which is affected by the degradation function $h(\phi)$, and $\psi^-_0(\mathbf{\varepsilon})$, which is not,
\begin{equation}
\label{eq:psi_schema}
    \psi(\phi,\mathbf{\varepsilon}) := \psi^+_0(\mathbf{\varepsilon})h(\phi) + \psi^-_0(\mathbf{\varepsilon}).
\end{equation}
(Note that the isotropic model in Eq.~\eqref{eq:isotropic_en} also fits this schema with $\psi^-_0=0$.)
In this work we consider the strain-spectral split of Miehe et al.\ \cite{miehe_phase_2010} and the volumetric-deviatoric split of Amor et al.\ \cite{amor_regularized_2009}, both formulated for an otherwise isotropic material.
These are two of the most widely studied tension/compression splits (see, e.g., Refs.~\cite{ambati_review_2015,wu_chapter_2020,bilgen_crack-driving_2019,freddi_regularized_2010,de_lorenzis_nucleation_2021,zhang_assessment_2022}).
The strain-spectral split has terms $\psi^+_0$ and $\psi^-_0$ of the form
\begin{equation}
\label{eq:strain_spectral_en}
    \psi^\pm_0(\mathbf{\varepsilon}) = \frac{1}{2} \lambda \left< \sum^d_{\alpha=1} \mathbf{\varepsilon}_\alpha \right>_\pm^2 +  \mu \sum^d_{\alpha=1} \left<  \mathbf{\varepsilon}_{\alpha} \right>_\pm^2,
\end{equation}
where $\mathbf{\varepsilon}_\alpha$ are the eigenvalues of the strain $\mathbf{\varepsilon}$ and the angle brackets $\langle \cdot \rangle_\pm$ denote ramp functions such that $\langle x \rangle_+=x$ for $x>0$,  $\langle x \rangle_-=x$ for $x<0$, and both functions are zero otherwise.
The volumetric-deviatoric split takes the form
\begin{equation}
\label{eq:vol_dev_en}
    \psi^+_0(\mathbf{\varepsilon}) = \frac{1}{2}K \left< \mathrm{tr}(\mathbf{\varepsilon}) \right>_+^2 +  \mu \sum_{i=1}^d \sum_{j=1}^d \left( \varepsilon_{ij} - \frac{1}{3}\delta_{ij} \mathrm{tr}(\mathbf{\varepsilon})  \right)^2,
\end{equation}
\[
\psi^-_0(\mathbf{\varepsilon}) = \frac{1}{2} K \left< \mathrm{tr}(\mathbf{\varepsilon}) \right>_-^2
\]
where $\delta_{ij}$ is the Kronecker delta and $K$ is the bulk modulus, $K = \lambda + 2\mu/3$.
As written in Eq.~\eqref{eq:vol_dev_en}, this formulation holds for 3D and 2-D cases such plane strain and plane stress that are obtained from 3D \cite{li_phase_2021}; Amor et al.\ \cite{amor_regularized_2009} additionally proposed a purely 2-D formulation that we will not consider here.

The total dissipated fracture energy $F_f$ is formulated to approximate its theoretical equivalent for a sharp crack,
\begin{equation}
    F_{f,\mathrm{sharp}}: = \int_\Gamma G_c \dif \mathcal H^{d-1},
\end{equation}
where $\Gamma$ is the set corresponding to a sharp crack, $\mathcal H^{d-1}$ is the $d-1$-dimensional Hausdorff measure (equivalent to length for $d=2$ or area for $d=3$ for sufficiently regular $\Gamma$), and $G_c$ is the critical energy release rate, the energy dissipated when $\Gamma$ is extended by a unit of $\mathcal H^{d-1}$ under equilibrium conditions.
Phase field models approximate $F_{f,\mathrm{sharp}}$ via an elliptic functional in $\phi$ \cite{ambrosio_approximation_1990,bourdin_numerical_2000,gerasimov_penalization_2019},
\begin{equation}
\label{eq:Ff}
    F_f[\phi] := \frac{G_c}{\ell c_w} \int_\Omega \left[ f(\phi) + \ell^2 |\nabla \phi|^2 \right] \dif \mathbf x,
\end{equation}
where $\ell$ is a length scale that determines the width of the diffuse crack, and $c_w$ is a constant that takes different values depending on the form of $f(\phi)$ to ensure that $F_f$ evaluates to $G_c$ for an ideal phase field crack with unit $\Gamma$.
The actual increment in $F_f$ corresponding to an unit increment in $\Gamma$ is usually larger than $G_c$ in practice, for example due to numerical error \cite{linse_convergence_2017,bleyer_dynamic_2017}.
For systems in which we can easily measure $\Gamma$, we denote this `true' energy release rate by $G=\dif F_f/\dif\mathcal H^{d-1}$.

We consider two forms for the local fracture energy density term $f(\phi)$ \cite{pham_gradient_2011,gerasimov_penalization_2019},
\begin{equation}
\label{eq:fracturelocal}
    \textrm{(AT1):} \;\; f(\phi)= \phi, \; c_w=8/3; \;\;\;\textrm{(AT2):} \;\; f(\phi) = \frac{1}{2} \phi^2,\; c_w = 2.
\end{equation}
In combination with the quadratic degradation function $h(\phi)=(\phi-1)^2$, these forms of $f(\phi)$ correspond to the AT1 and AT2 models considered in Refs.\ \cite{tanne_crack_2018,alessi_comparison_2018}.
The `AT' designation refers to Ambrosio and Tortorelli \cite{ambrosio_approximation_1990}, who provided a method to prove $\Gamma$-convergence of $F_f$ to $F_{f,\mathrm{sharp}}$ in the limit $\ell \to 0$.
The AT2 model corresponds to the original phase field fracture model proposed by Bourdin et al.\ \cite{bourdin_numerical_2000}, while AT1 was proposed subsequently by Pham et al.\ \cite{pham_gradient_2011}.
The AT1 and AT2 models result in different optimal profiles of $\phi(x)$ for a 1-D crack \cite{miehe_thermodynamically_2010,pham_gradient_2011}:
\begin{equation}
    \label{eq:AT1_analytical}
    \mathrm{(AT1):}\;\;\;\phi(x) = \left(1- \frac{|x-x_0|}{2\ell}\right)^2,
\end{equation}
\begin{equation}
    \label{eq:AT2_analytical}
    \mathrm{(AT2):}\;\;\;\phi(x) = \exp{\frac{-|x-x_0|}{\ell}},
\end{equation}
where $x_0$ denotes the center of the crack.

During simulations with the AT2 model, the phase field increases as soon as $\psi^+_0$ becomes non-zero, which prevents truly elastic behavior and leads to delocalized evolution of $\phi$ far from the eventual crack \cite{pham_gradient_2011}.
This delocalized evolution results in a worse description of crack nucleation in systems that lack a strongly singular stress concentration compared to the AT1 model \cite{tanne_crack_2018}, which retains a linear elastic response until the onset of fracture.
The principal disadvantage of the AT1 model is that it is ill posed unless a constraint is imposed on $\phi$ throughout the entire domain: either the irreversibility constraint must be enforced throughout the entire domain or another constraint (e.g., $\phi \ge 0$) must be added where the irreversibility constraint is not enforced.
The AT2 model has no such requirement due to $f(\phi)$ being strictly convex.

Enforcing irreversible evolution of $\phi$ in the entire domain has been found to negatively affect $\Gamma$-convergence of $F_f$ with the AT2 model due to the delocalized evolution of $\phi$ prior to fracture \cite{linse_convergence_2017}.
Thus, works with the AT2 model often limit the irreversibility constraint to a crack set of points with $\phi$ greater than some threshold value \cite{bourdin_numerical_2000,gerasimov_penalization_2019}.
To provide consistent notation between these constraints, we define two variants of a constraining field $\phi_\mathrm{con.}(\phi)$,
\begin{equation}
\label{eq:constraint_field}
   \textrm{(crack-set):}\;\; \phi_\mathrm{con.}(\phi) = 
\begin{cases}
    \phi& \text{if}\; \phi \geq 0.9,\\
    0 & \text{otherwise}
\end{cases},\;\;\;\;
\textrm{(damage):}\;\; \phi_\mathrm{con.}(\phi) = \phi.
\end{equation}
where the crack set has been approximated as the set of points where $\phi(\mathbf x)>0.9$ and the `damage' name refers to the prevalence of irreversibility in the entire domain in interpretations of phase field fracture as a damage model \cite{pham_gradient_2011,de_borst_gradient_2016}.
The irreversibility constraint can then be written as $\phi - \phi_\mathrm{con.} \ge 0$, where $\phi_\mathrm{con.}$ is obtained from Eq.~\ref{eq:constraint_field} based on a previous iterate for $\phi$.
The choice of previous iterate differs between evolution methods.

We may now write the overall energy functional for the phase field fracture model as
\begin{equation}
\label{eq:functional}
    F[\phi, \mathbf u] = \int_\Omega \psi\left(\phi,\mathbf{\varepsilon}(\nabla \mathbf u)\right) + \frac{G_c}{c_w \ell}\left( f(\phi) + \ell^2 |\nabla \phi|^2 \right) \dif\mathbf x,
\end{equation}
This work will focus on the choices of $f(\phi)$ in Eq.\ \eqref{eq:fracturelocal} and the choices of $\psi(\phi,\mathbf{\varepsilon})$ described in Eqs.\ \eqref{eq:isotropic_en}-\eqref{eq:vol_dev_en}.
This selection of formulations is intended to represent the simplest and most commonly used formulations for quasi-static brittle fracture, and is not comprehensive.
See for example Refs.\ \cite{pham_gradient_2011,wu_unified_2017,wu_chapter_2020,wu_length_2018} for alternative forms of the local fracture energy density $f(\phi)$ and degradation function $h(\phi)$, Ref.\ \cite{borden_higher-order_2014} for a form of $F_f$ incorporating the Laplacian of $\phi$, and Refs.\ \cite{bilgen_crack-driving_2019, de_lorenzis_nucleation_2021,zhang_assessment_2022} for alternative decompositions of the elastic energy density $\psi(\phi,\mathbf{\varepsilon})$.

Instead of $F$ itself, evolution methods use $F_\phi$ and $F_{\mathbf u}$, respectively the variational derivatives of $F$ with respect to $\phi$ and $\mathbf u$.
For $F$ as written in Eq.\ \eqref{eq:functional}, these variational derivatives are
\begin{equation}
    \label{eq:F_phi}
    F_\phi = h'(\phi)\psi^+_0(\mathbf{\varepsilon}) + \frac{G_c}{c_w \ell}\left( f'(\phi) - \ell^2 \nabla^2 \phi \right)
\end{equation}
\begin{equation}
    \label{eq:F_strain}
    F_\mathbf{u} = -\nabla \cdot \mathbf{\sigma}, \;\;
    \mathbf{\sigma} = \frac{\partial \psi(\mathbf{\varepsilon},\phi)}{\partial \mathbf{\varepsilon}}.
\end{equation}
where the symmetrization operator $\partial \varepsilon/\partial \nabla \mathbf{u}$ has no effect for the choices of $\psi(\mathbf{\varepsilon},\phi)$ considered here.
Like the energy density itself, the stress can be expressed as a splitting of two terms modified by the degradation function $h(\phi)$,
\begin{equation}
    \mathbf{\sigma} = \frac{\partial \psi^+_0}{\partial \mathbf{\varepsilon}}h(\phi) +\frac{\partial \psi^-_0}{\partial \mathbf{\varepsilon}} = \mathbf{\sigma}_0^+ h(\phi) + \mathbf{\sigma}_0^-.
\end{equation}
The stress decompositions for the isotropic, strain-spectral, and volumetric-deviatoric models are then
\begin{equation}
\label{eq:stress_iso}
    \textrm{(isotropic):}\;\; \mathbf{\sigma}^+_0 = \lambda \mathbf I \mathrm{tr}(\mathbf{\varepsilon}) + 2\mu \varepsilon_{ij},\;\; \mathbf{\sigma}^-_0 = \mathbf 0
\end{equation}
\begin{equation}
\label{eq:stress_spectral}
    \textrm{(strain-spectral):}\;\; \mathbf{\sigma}^\pm_0 = \sum_{\alpha=1}^d \left( \lambda \left< \mathrm{tr}(\mathbf{\varepsilon})\right>_\pm + 2\mu \left< \mathbf{\varepsilon}_\alpha \right>_\pm \right) \mathbf{n}^\alpha \otimes \mathbf{n}^\alpha,
\end{equation}
\begin{equation}
\label{eq:stress_voldev}
    \textrm{(volumetric-deviatoric):}\;\;\mathbf{\sigma}^+_0 = \frac{1}{3} K \mathbf I \left< \mathrm{tr}(\mathbf{\varepsilon})\right>_+ + 2\mu \left[\mathbf{\varepsilon} - \frac{1}{3} \mathbf I \left<\mathrm{tr}(\mathbf{\varepsilon})\right>_+ \right],\;\; \mathbf{\sigma}^-_0 =\frac{1}{3} K\mathbf I \left< \mathrm{tr}(\mathbf{\varepsilon})\right>_-,
\end{equation}
where $\mathbf{n}^\alpha$ is the $\alpha$-th eigenvector of $\mathbf{\varepsilon}$, $\otimes$ denotes the outer product, $\mathbf 0$ is the $d\times d$ matrix with all entries equal to zero, and $\mathbf I$ is the $d\times d$ identity matrix.

Under a variety of circumstances, it can be convenient to change the terms $\partial \psi/\partial \phi$ and $\partial \psi/\partial \mathbf{\varepsilon}$ in Eqs.~\eqref{eq:F_phi} and \eqref{eq:F_strain}, respectively, such that they no longer represent derivatives of the same energy density $\psi$.
The term $\partial \psi/\partial \phi$ has become known as the crack driving force \cite{bilgen_crack-driving_2019,kumar_revisiting_2020}.
The different forms of the stress affect the mechanical response of the crack and other regions with non-zero $\phi$.
For this reason, we refer to forms of $\partial \psi/\partial \mathbf{\varepsilon}$ as contact models, even if they fail to reproduce realistic contact physics \cite{amor_regularized_2009,freddi_regularized_2010,zhang_assessment_2022}.
The earliest example of a non-variational phase field fracture model may be the use of a history function in place of $\psi_0^+$ in the crack driving force in order to satisfy a damage-type irreversibility condition \cite{miehe_phase_2010,gerasimov_penalization_2019}.
Ambati et al.\ \cite{ambati_review_2015} proposed using the crack driving force from the strain-spectral split, Eq.\ \eqref{eq:strain_spectral_en}, with the stress-free contact model from the isotropic formulation, Eq.\ \eqref{eq:isotropic_en}, to save on computational effort.
Other works have proposed non-variational forms of the crack driving force to better approximate experimental strength surfaces \cite{wu_unified_2017,kumar_revisiting_2020} and crack paths \cite{bilgen_crack-driving_2019}.
In this work, we will only consider crack driving forces and contact models derived from the energy densities in Eqs.\ \eqref{eq:isotropic_en}-\eqref{eq:vol_dev_en}, but we will consider non-variational combinations of crack driving forces and contact models.

\subsection{Evolution methods} \label{sec:bg_evolution}
We can consider three main types of models for the evolution of $\phi$ during quasi-static brittle fracture \cite{ambati_review_2015,wu_chapter_2020}: minimization \cite{bourdin_numerical_2000}, time-dependent evolution \cite{karma_phase-field_2001,miehe_phase_2010, miehe_thermodynamically_2010}, and near-equilibrium or path-following evolution \cite{vignollet_phase-field_2014,may_new_2016,singh_fracture-controlled_2016}.

In the minimization approach, the functional $F[\phi, \mathbf{\varepsilon}]$ is minimized with respect to $\phi$ and $\mathbf u$ \cite{bourdin_numerical_2000},
\begin{equation}
\label{eq:minimization}
   \phi, \mathbf u =   \mathrm{arg} \min_{\phi',\mathbf u'} F[\phi',\mathbf u'].
\end{equation}
This minimization is complicated by the non-convexity of the $\psi(\phi,\mathbf{\varepsilon})$ term in $F$, which, depending on the method used, may result in non-convergence \cite{gerasimov_line_2016,heister_primal-dual_2015,farrell_linear_2017} or convergence to a local rather than a global minimizer \cite{bourdin_numerical_2007}.
Bourdin \cite{bourdin_numerical_2007} discusses the issue of global vs.\ local minimizers in depth and provides a backtracking method for finding global minimizers.
However, in subsequent literature it has been common to accept the local minimizers resulting from a particular optimization algorithm as the solution \cite{heister_primal-dual_2015,ambati_review_2015,gerasimov_line_2016,farrell_linear_2017,wick_modified_2017,gerasimov_penalization_2019,ernesti_fast_2020}, although finding an ensemble of local minimizers has also been proposed \cite{gerasimov_stochastic_2020}.
Finding a local minimizer amounts to finding $\phi$ and $\mathbf u$ that satisfy the Karoush-Kuhn-Tucker optimality conditions \cite{pham_gradient_2011}, namely the stationary condition for $\mathbf u$,
\begin{equation}
\label{eq:stationarity_u}
   F_\mathbf{u} = 0,
\end{equation}
the stationarity and dual feasibility conditions for $\phi$,
\begin{equation}
\label{eq:stationarity_phi}
   F_\phi \ge 0,
\end{equation}
the irreversibility condition on $\phi$ (primal feasibility),
\begin{equation}
\label{eq:irreversibility}
   \phi - \phi_\mathrm{con.} \ge 0
\end{equation}
and the complementary slackness condition for $\phi$,
\begin{equation}
\label{eq:slackness}
    F_\phi \left(\phi - \phi_\mathrm{con.} \right) = 0,
\end{equation}
where $\phi_\mathrm{con.}$ is based on the previous minimization result.
Minimization allows brutal fracture where a minimization step results in a discontinuous change in $\phi$, often corresponding to sudden propagation of a crack through the domain \cite{francfort_revisiting_1998,bourdin_numerical_2000,bourdin_numerical_2007,bourdin_variational_2008}.
For such cases, the irreversibility constraint plays a much smaller role compared to other evolution methods.
Typical solution methods for minimization are Newton-based monolithic schemes \cite{heister_primal-dual_2015,gerasimov_line_2016,farrell_linear_2017,wick_modified_2017,gerasimov_penalization_2019} and alternating minimization (AM), in which the solver alternates between solving Eq.\ \eqref{eq:stationarity_u} with $\phi$ held constant and Eqs.\ \eqref{eq:stationarity_phi}-\eqref{eq:slackness} with $\mathbf u$ held constant until a convergence criterion is reached \cite{bourdin_numerical_2000,bourdin_numerical_2007,hossain_effective_2014,ambati_review_2015,farrell_linear_2017}.

Time-dependent evolution, the second type of evolution method, can be interpreted either as a viscous regularization of the minimization method \cite{miehe_thermodynamically_2010} or a Ginzburg-Landau-type gradient flow \cite{lazzaroni_model_2011,kuhn_continuum_2010},
\begin{equation}
\label{eq:time_evolution}
    \eta \frac{\partial \phi}{\partial t} \ge - F_\phi,
\end{equation}
where $\eta$ is a viscosity parameter.
The displacement field is governed by Eq.\ \eqref{eq:stationarity_u}, the irreversibility condition Eq.\ \eqref{eq:irreversibility} is applied with $\phi_\mathrm{con.}$ based on the previous time step, and the equivalent of the complementary slackness condition, Eq.\ \eqref{eq:slackness}, is
\begin{equation}
\label{eq:slack_time}
    \left( \eta \frac{\partial \phi}{\partial t} + F_\phi \right)\left( \phi - \phi_\mathrm{con.} \right) = 0.
\end{equation}
Unlike minimization-based methods, the time-dependent method regularizes brutal fracture: in the limit of continuous time evolution, the time-dependent method results in `progressive' fracture where $\phi$ changes continuously between steps \cite{lazzaroni_model_2011,bourdin_variational_2008}.
Like the choice of a specific algorithm in the minimization method, the time-dependent method evolves along a specific pathway for energy dissipation and crack growth during fracture \cite{bourdin_variational_2008}.
However, even if the minimization method is applied with an iterative algorithm similar in form to Eq.\ \eqref{eq:time_evolution}, it would still be mathematically distinct from the time-dependent method because it enforces irreversibility based on the initial state of the minimization algorithm, rather than the previous update.
In this sense, the staggered method proposed by Miehe et al.\ \cite{miehe_phase_2010}, in which the irreversibility condition is updated after a single iteration of the alternating minimization algorithm, can be interpreted as a time-dependent method in the limit of zero viscosity, $\eta \to 0$.
We note that the time-continuous crack path will only be affected by $\eta$ if there is another source of time dependence in the system (e.g., in the loading conditions).
If there is no other time dependence, then $\eta$ can be combined with the discrete time step $\Delta t$ into a numerical parameter $\Delta t/\eta$, where low $\Delta t/\eta$ corresponds to less evolution per step.

The third type of evolution model is what we call near-equililbrium methods.
The reason fracture simulations do not tend to remain near equilibrium is illustrated by linear elastic fracture mechanics, which predicts that for a crack in Mode I loading, the energy release rate $G$ increases linearly as a function of crack length \cite{zehnder_fracture_2012,rice_mathematical_1968}.
Thus, once a crack starts to grow, $G$ will continue to increase beyond $G_c$, drawing the system further from equilibrium.
Similar behavior is widely seen in mechanical systems with strain-softening properties, and is referred to as snap-back \cite{de_borst_computation_1987,singh_fracture-controlled_2016} due to the simultaneous decreases in stress and strain on an equilibrium stress-strain plot.
If the reduction in loading did not occur, the system would be far from equilibrium in an overstressed state.
(One could also refer to this state as overstrained, but `overstrained' is associated with plasticity moreso than fracture \cite{vincent_mechanics_1992}).
Overstress is known to affect crack morphology and dissipated energy in experiments \cite{scheibert_brittle-quasibrittle_2010} and simulations of dynamic fracture \cite{bleyer_dynamic_2017}.
We define near-equilibrium methods as methods where the loading conditions are adapted during evolution to remain near equilibrium, leading to progressive crack growth in which the irreversibility condition is applied between steps.

The main category of near-equilibrium method is known as path-following or arc-length control methods.
In these methods, $F$ is minimized subject to a constraint that a quantity that increases monotonically during fracture (e.g., dissipated energy \cite{gutierrez_energy_2004,vignollet_phase-field_2014,may_new_2016} or crack set measure \cite{singh_fracture-controlled_2016}) must increase by a fixed amount $\Delta \tau$.
To provide the additional degree of freedom to satisfy this constraint, the applied boundary conditions are allowed to vary, typically via a single scaling parameter.
The augmented system, composed of the original constrained minimization problem plus the path-following constraint, models progressive fracture along a path that is as close as possible to satisfying the equilibrium conditions, Eqs.\ \eqref{eq:stationarity_u}-\eqref{eq:slackness}, given the discrete increment in the control parameter $\tau$.

Near-equilibrium behavior can be recovered in other evolution methods through specific choices of geometry and/or boundary conditions.
For instance, Hossain et al.\ \cite{hossain_effective_2014} proposed a `surfing' boundary condition in which crack growth via any evolution method is self-limiting.
These surfing boundary conditions consist of Dirichlet conditions on the displacements based on the LEFM solution for a crack tip at a given location; propagation is then driven by increasing the magnitude of the displacements and/or translating the imposed crack tip location.
A large pre-existing crack normal to the loading direction will also limit snap-back by limiting the amount by which crack propagation can increase $G$ \cite{zehnder_fracture_2012}.

\section{Methods}

\subsection{Sub-problem solution methods}
In this work, we consider numerical approaches in which the phase field and elasticity sub-problems are weakly coupled in that separate linear-algebraic problems are solved for each sub-problem.
This can simplify implementation by allowing the use of standalone mechanics and/or phase field codes developed for other problems, albeit usually at the cost of performance compared to `monolithic' methods that solve both fields simultaneously \cite{gerasimov_line_2016}.
In our case, weak coupling makes it easier to apply FFT-based preconditioning for the elasticity problem, which improves computational performance and enables scalability.

We solve the elasticity sub-problem via a Fourier Galerkin scheme and the phase field sub-problem via a Fourier collocation scheme.
Both of these schemes employ the same representations of the fields in real and Fourier space.
In the Fourier Galerkin scheme, trigonometric polynomials are used as test functions and a quadrature rule is applied to solve the equations in weak form.
In the collocation scheme, a trigonometric projection operator is applied to the governing equations resulting in an expression for the strong form of the governing equations/inequalities at each grid point \cite{saranen_periodic_2002}.

Following \cite{zeman_finite_2017}, we consider a 2-D domain $\Omega$ centered at the origin with lengths $L_x$ and $L_y$ in the $x$ and $y$ directions: $\Omega = [-L_x/2,L_x/2]\times [-L_y/2,L_y/2] \subset \mathbb R^2$, with area $|\Omega| = L_x L_y$.
To discretize this domain, we define a regular 2-D grid.
We denote the size of the grid by the vector $\mathbf N = (N_x, N_y) \in \mathbb N^2$, where $N_x$ and $N_y$ are the numbers of points in each direction and $|\mathbf N|=N_x N_y$ is the total number of grid points.
We can then define a set of grid point indices as
\begin{equation}
    \mathbf Z_N^2 = \left\{\mathbf k = (k_x,k_y) \in \mathbb Z^2  : \frac{-N_x}{2} < k_x < \frac{N_x}{2} , \frac{-N_y}{2} < k_y < \frac{N_y}{2} \right\}
\end{equation}
The vector of coordinates $\mathbf x$ for the grid point corresponding to index $\mathbf k$ is
\begin{equation}
    \mathbf x^\mathbf{k} = \left(\frac{k_x L_x}{N_x}, \frac{k_y L_y}{N_y} \right).
\end{equation}
Likewise, the wavevector $\mathbf q$ corresponding to index $\mathbf k$ is
\begin{equation}
    \mathbf q^\mathbf{k} = \left(\frac{k_x }{L_x}, \frac{k_y}{L_y} \right).
\end{equation}
Now we define the space of trigonometric polynomials,
\begin{equation}
    \mathcal T_N = \left\{ \sum_{\mathbf k \in \mathbb Z_N^2} c_\mathbf{k} e^{2\pi i \mathbf q^\mathbf{k} \cdot \mathbf x}: c_\mathbf{k} \in \mathbb C, \mathbf k \in \mathbb Z_N^2 \right\}.
\end{equation}
For a function $v \in \mathcal T_N$, the coefficents $\hat v_N$ of its trigonometric polynomial are determined by its discrete Fourier transform $\mathcal F_N$,
\begin{equation}
    \mathcal F_N v (\mathbf x^\mathbf{j}) = \hat v (\mathbf q^\mathbf{k}) = \frac{1}{|\mathbf N|} \sum_{\mathbf j \in \mathbb Z_N^2} v (\mathbf x^\mathbf{j}) \exp \left(- 2\pi i \mathbf q^{\mathbf k} \cdot \mathbf x^\mathbf{j} \right), \; \;\; (\mathbf j, \mathbf k \in \mathbb Z_N^2)
\end{equation}
(The circumflex $\hat \cdot$ is used hereafter to indicate the Fourier coefficients of a real-space field or operator.)
Likewise, values of $v_N$ at grid points $\mathbf x^\mathbf{k}$ can be obtained by the inverse transform $\mathcal F_N^{-1}$,
\begin{equation}
    \mathcal F_N^{-1} \hat v (\mathbf q^\mathbf{k}) = v (\mathbf x^\mathbf{j}) = \sum_{\mathbf k \in \mathbb Z_N^2} \hat v (\mathbf q^\mathbf{k}) \exp \left( 2\pi i \mathbf q^{\mathbf k} \cdot \mathbf x^\mathbf{j} \right), \; \;\; (\mathbf j, \mathbf k \in \mathbb Z_N^2).
\end{equation}
An additional property, relevant for the Fourier Galerkin scheme, is that an inner product of functions $v,w \in \mathcal T_N$ over $\Omega$ is exactly equal to the integration of their product by the trapezoidal method,
\begin{equation}
\label{eq:trapezoidal}
    \int_\Omega v(\mathbf x) w(\mathbf x) \dif\mathbf x = \frac{|\Omega|}{|\mathbf N|} \sum_{\mathbf k \in \mathbb Z_N^2} v(\mathbf x^\mathbf{k}) w(\mathbf x^\mathbf{k}),
\end{equation}
when the numbers of grid points in each direction, $N_x$ and $N_y$, are both odd.
For this reason, we only consider odd $N_x$ and $N_y$ here.
C.f. Refs.\ \cite{zeman_finite_2017,leute_elimination_2022,ladecky_optimal_2022} for the general case and additional details regarding this property.

\subsubsection{Elasticity sub-problem}
The elasticity sub-problem is solved by a Fourier Galerkin scheme with the strain field $\mathbf{\varepsilon}$ as the principal unknown.
The strain field is considered to be $\Omega$-periodic, and it is decomposed as $\mathbf{\varepsilon} = \mathbf{\bar \varepsilon} + \mathbf{\varepsilon}^*$ into a constant term $\mathbf{\bar \varepsilon} = \frac{1}{|\Omega|}\int_\Omega \mathbf{\varepsilon} \dif\mathbf x$ and a polarization term $\mathbf{\varepsilon}^*(\mathbf x)$ that is spatially varying and has zero mean, $\int_\Omega \mathbf{\varepsilon}^* \dif\mathbf x=0$.
Loading is applied by setting $\mathbf{\bar \varepsilon}$, leaving $\mathbf{\varepsilon}^*$ to be determined by the Fourier Galerkin scheme.
The conditions to be satisfied are mechanical equilibrium,
\begin{equation}
    \label{eq:elasticity}
    \nabla \cdot \mathbf{\sigma}=0 \qquad \text{(see also Eq.\ \eqref{eq:F_strain})},
\end{equation}
and compatibility of the spatially varying strain field $\mathbf{\varepsilon}^*$: $\mathbf{\varepsilon}^*=\nabla_s \mathbf u^*$ for some $\Omega$-periodic displacement vector $\mathbf u^*$.
Implicit in this definition of the compatibility condition is the fact that we are using a small-strain formulation of elasticity, which is typical for phase field fracture models.

The first step towards deriving the Fourier Galerkin scheme is the statement of the weak form of Eq.\ \eqref{eq:elasticity},
\begin{equation}
\label{eq:elasticity_weak}
    \int_\Omega \delta \mathbf{\varepsilon}^*:\mathbf{\sigma} d\mathbf x = 0,
\end{equation}
where $\delta \mathbf{\varepsilon}^*$ denotes a test function from within the space of compatible tensor fields and the stress $\mathbf{\sigma}$ is expressed in terms of $\mathbf{\varepsilon}$ in Eqs.~\eqref{eq:F_strain}-\ref{eq:stress_voldev}.
When $\delta \mathbf{\varepsilon}^*$ and $\mathbf{\sigma}$ are both members of $\mathcal T^{2\times 2}_N$, the space of rank-2 tensor fields with components in $\mathcal T_N$, Eq.\ \eqref{eq:trapezoidal} implies that the weak form in Eq.\ \eqref{eq:elasticity_weak} is equivalent to the following discrete integration:
\begin{equation}
\label{eq:elasticity_weak_discrete}
    \frac{|\Omega|}{|\mathbf N|} \sum_{\mathbf k \in \mathbb Z_N^2} \delta \mathbf{\varepsilon}^*(\mathbf x^\mathbf{k}) :\mathbf{\sigma}(\mathbf x^\mathbf{k}) = 0.
\end{equation}
Since it is not known a priori if an arbitrary test function $\zeta \in \mathcal T^{2\times 2}_N$ is compatible, we decompose our compatible test function $\delta \mathbf{\varepsilon}^*$ into the convolution of an arbitrary test function $\zeta$ with an operator $\mathbf G$ that projects it into the subspace of $\mathcal T^{2\times 2}_N$ consisting of compatible strain fields,
\begin{equation}
    \delta \mathbf{\varepsilon}^* = \int_\Omega \mathbf G(\mathbf x-\mathbf x') \zeta (\mathbf x)d\mathbf x'.
\end{equation}
This convolution is symmetric and sparse in Fourier space as the Fourier-space operator $\hat G$ is block diagonal (see e.g., Refs.\ \cite{milton_variational_1988, zeman_finite_2017, leute_elimination_2022, Ladecky2022-kl} for its precise form).
Now, taking the discrete Fourier transform of Eq.\ \eqref{eq:elasticity_weak_discrete} and substituting $\hat {\delta \mathbf{\varepsilon}^*} = \hat \zeta : \hat G$, we have the following discretized weak form,
\begin{equation}
    \frac{|\Omega|}{|\mathbf N|^2} \sum_{\mathbf k \in \mathbb Z_N^2} \hat \zeta^\mathbf{k} : \mathbf{ \hat G}^\mathbf{k}: \mathbf{\hat \sigma}^\mathbf{k} = 0.
\end{equation}
which results in the nodal equilibrium equations
\begin{equation}
\label{eq:elasticity_nodal}
   \mathbf{\hat G}^\mathbf{k}:\hat \sigma^\mathbf{k} = 0.
\end{equation}

The system of nodal equations for $\mathbf{\varepsilon}^*$ may be non-linear, and thus we apply Newton's method to solve for $\mathbf{\varepsilon}^{*}$,
\begin{equation}
\label{eq:mech_newton}
    \mathbf{\varepsilon}^*_{m+1} = \mathbf{\varepsilon}^*_m + \theta_m,
\end{equation}
where the Newton update $\theta_m$ at step $m$ is obtained by using conjugate gradients (CG) to solve
\begin{equation}
\label{eq:elasticity_newton_linear}
    \mathbf{\hat G} :\widehat{\mathbf C_{m}:\theta_m}= \mathbf{\hat G}:\mathbf{\hat \sigma}_m,
\end{equation}
where $\mathbf C_m$ is the stiffness tensor $\mathbf C = \partial \mathbf{\sigma}/\partial \mathbf{\varepsilon}^*$ at step $m$.
The stress $\mathbf{\hat \sigma}$ and the matrix product $\mathbf C_m:\theta_m$ are computed for each real-space grid point, taking into account any spatial differences in material properties
Then, their FFTs are taken in order to apply the projection operator $\mathbf G$ in Fourier space.
This numerical method is highly efficient due to the sparsity of the linear operations in real space (calculation of the stresses) and Fourier space (application of the projection operator) and the efficiency of the only dense operation, the FFT \cite{zeman_finite_2017}.
It also benefits from almost optimal conditioning~\cite{gergelits_laplacian_2019,ladecky_optimal_2022}.

Overall, the mechanics sub-problem only differs between different mechanics models; it is unaffected by the choices of evolution method or fracture energy formulation $F_f$ considered here.
For the three mechanics models we consider (isotropic, strain-spectral split, and volumetric-deviatoric split), forms of $\mathbf C$ are available in literature \cite{li_phase_2021}.
In all three cases, the nodal equations, Eq.\ \eqref{eq:elasticity_nodal}, are solved to either a relative or absolute tolerance in the $L_\infty$ norm, $||f||_\infty = \max f$.
Since $\mathbf C$ is linear in the isotropic model, the Newton iteration in Eq.\ \eqref{eq:mech_newton} is terminated after a single step in which the linear problem in Eq.\ \eqref{eq:elasticity_newton_linear} is solved via CG to the final desired tolerance.
For the strain-spectral split and volumetric-deviatoric split, a lower relative tolerance is used for CG solves for the Newton updates.
Our simulations employ a plane-stress formulation in which the out-of-plane strains are not represented explicitly, meaning that we work in a reduced $2\times 2$ representation of the strain.

\subsubsection{Phase field sub-problem}
The phase field $\phi$ is discretized in space by a Fourier collocation scheme employing the same space of basis functions $\mathcal T_N$ that the Galerkin scheme used for the components of $\epsilon^*$.
In this scheme, we solve the strong forms of the equations/inequalities \eqref{eq:stationarity_phi}-\eqref{eq:slackness} at each grid point $\mathbf x_k \in \mathbb Z^2_N$.
The only term in these expressions that requires information from other grid points is the Laplacian $\nabla^2\phi$ in $F_\phi$.
To evaluate the Laplacian in this discretization, we define the collocation Laplacian $\nabla^2_N$,
\begin{equation}
    \nabla^2_N g =-\mathcal F_N^{-1} || 2\pi\mathbf q||^2   \mathcal F_N g, \;\; g \in \mathcal T_N.
\end{equation}
In addition to the spatial discretization, discretization in time is required for time-dependent evolution methods for the phase field.
We implement the time-dependent evolution via a backwards Euler scheme,
\begin{equation}
\label{eq:phase_field_euler}
    \frac{\eta}{\Delta t}\left( \phi_{n} - \phi_{n-1} \right) \ge -F_\phi(\phi_n),
\end{equation}
where $\Delta t$ is the time increment and $n$ is the index of the time increment.
Time-independent formulations are considered by taking $\eta/\Delta t=0$.
Inequality \eqref{eq:phase_field_euler} is linear in $\phi_n$ for the choices of $f(\phi)$ and $h(\phi)$ considered here.
To formulate the linear unconstrained problem in a general way, we consider a Newton-type update $v_r = \phi_r - \phi_{r-1}$,
\begin{equation}
\label{eq:phase_field_update}
    J v_r \ge -F_\phi(\phi_{r-1}),
\end{equation}
where $J$ is the Jacobian matrix of Eq.\ \eqref{eq:phase_field_euler},
\begin{equation}
    J = \frac{\eta}{\Delta t} + h''(\phi)\psi^+ - \frac{G_c}{c_w \ell} \left[ f''(\phi) - \ell^2 \nabla^2_N \right],
\end{equation}
in which $h''(\phi)=2$ and $f''(\phi)$ is equal to zero for AT1 and $1/2$ for AT2.
The irreversibility constraint on the update $v_r$ is formulated as $v_r \ge v_\mathrm{con.}$, where $v_\mathrm{con.} = \phi_{r-1} - \phi_\mathrm{con.}(\phi_{r-1})$, with $\phi_\mathrm{con.}$ defined in Eq.~\ref{eq:constraint_field}.

We solve Eq.\ \eqref{eq:phase_field_update} subject to the irreversibility constraint and slackness condition using a bound-constrained conjugate gradients (BCCG) algorithm, specifically the enhanced BCCG(K) algorithm introduced by Vollebregt \cite{vollebregt_bound-constrained_2014}.
Convergence of this algorithm is not in general guaranteed; Vollebregt conjectured that it converges for non-negative matrices, but this is not the case for $J$ due to the Laplacian operator $\nabla^2_N$.
Nevertheless, we find that it convergences to the desired numerical tolerance in all cases explored here.
In Algorithm \ref{alg:BCCG} below, we provide a concise statement of the BCCG algorithm as implemented in our code.

\begin{algorithm}[H]
\SetAlgoLined
Initialize solution vector $v^0$ (e.g., $v^0=b$) and set $m=1$\\
Set $v^0: = v_\mathrm{con.}$ where $v^0 < v_\mathrm{con.}$\\
$\displaystyle r^0 := Jv^0-b$\\
Set $r^0 := 0$ where both $v^0 = v_\mathrm{con.}$ and $r^0 > 0$\\
$\displaystyle p^0 := - r^0$\\
\While{$||r^{m-1}||_2 < \mathrm{Tol}_\mathrm{PF}$}{
    $\displaystyle \alpha:= \frac{r^{m-1}\cdot p^{m-1}}{p^{m-1}\cdot J p^{m-1}}$\\
    $\displaystyle v^{m} := v^{m-1} + \alpha p^{m-1}$\\
    Set $v^m: = v_\mathrm{con.}$ where $v^m < v_\mathrm{con.}$\\
    $\displaystyle r^m := Jv^m-b$\\
    Set $r^m := 0$ where both $v^m = v_\mathrm{con.}$ and $r^m > 0$\\
    $\displaystyle \beta:=\frac{r^m\cdot(r^m-r^{m-1})}{\alpha p^{m-1}\cdot J p^{m-1}}$\\
    $\displaystyle p^m := -r^m + \beta p^{m-1}$\\
    Set $p^m := 0$ where both $v^m = v_\mathrm{con.}$ and $r^m > 0$\\
    $m:=m+1$\\
 }
 \caption{Bound-constrained CG algorithm \label{alg:BCCG}}
\end{algorithm}

Notation in Algorithm \ref{alg:BCCG} has been simplified from Eq.\ \eqref{eq:phase_field_update}: we have dropped the time step/outer solver index $n$ from $v$ and we denote the RHS by $b=-F_\phi(\phi_{n-1})$.
The definition of the active set (points where both $v^m = v_\mathrm{con.}$ and $r^m > 0$) makes use of the fact that the complementary slackness condition can be written in terms of $v$ and the residual $r=Jv-b$ as $(v-v_\mathrm{con.})r = 0$.
The notation in Algorithm \ref{alg:BCCG} interprets $v$, $r$, $p$, $Jv$, and $Jp$ as vectors with the same length (i.e., $N_xN_y$), such that $r\cdot p$ is the conventional inner product and $||r||_2$ is the $\ell_2$ norm.
The matrix $J$ is never represented explicitly, as only the matrix-vector products $Jv$ and $Jp$ are used.
These matrix-vector products are the most computationally intensive steps in Algorithm \ref{alg:BCCG} because the collocation Laplacian requires fast Fourier transforms that take $\mathcal O(|N|\log |N|)$ time.

\subsection{Evolution Algorithms}
In this sub-section, we describe our implementations of the evolution methods from Section \ref{sec:bg_evolution}.
The previous sub-section described separate sub-problems for determining the strain field given an applied average strain $\mathbf{\bar \varepsilon}$ and $\phi$ and for determining $\phi$ given $\mathbf{\varepsilon}$, the constraining field $\phi_\mathrm{con.}$, and the time step $\Delta t$.
Each sub-problem is converged to a relative or absolute tolerance based on the $\ell_2$ norm of the residual.
The evolution algorithms integrate these sub-problem solvers with methods that control or adapt $\mathbf{\bar \varepsilon}$ and $\Delta t$.

\subsubsection{Alternating miminization}

For our minimization approach, we employ the alternating minimization algorithm (Algorithm \ref{alg:alternating-minimization}), in which the elasticity and phase field problems are solved separately one after the other.
The system is solved to convergence for each strain increment, and the converged phase field for the previous strain increment is used for the irreversibility constraint of the current strain increment.
The algorithm consists of an outer loop (index $s$) where $\mathbf{\bar \varepsilon}$ is updated by a tensor-valued increment $\Delta \mathbf{\bar \varepsilon}$ and an inner loop (index $n$) for the iterative minimization itself.
The inner/minimization loop is considered converged when the difference in $\phi$ between consecutive inner iterations is less than a tolerance $\mathrm{Tol}_\mathrm{AM}$, $||\phi_n-\phi_{n-1}||_1 < \mathrm{Tol}_\mathrm{AM}$, where $||\cdot||_1$ is the $L^1$ norm, $||f||_1=\frac{|\Omega|}{|\mathbf N|} \sum_{\mathbf k \in \mathbb Z_N^2} f(\mathbf x^\mathbf{k})$.
For the outer loop, the maximum number of strain steps $s_\mathrm{max.}$ is usually set such that a stiffness-based termination criterion is triggered first.
This stiffness-based criterion, also used in the other evolution methods, is triggered when a measure of stiffness $C$, calculated as the ratio between the largest components of the average stress $\mathbf{\bar \sigma}$ and the average strain $\mathbf{\bar \varepsilon}$, falls below a reference value $\bar C_\mathrm{broken}$ that is intended to represent the crack passing through most or all of the domain (e.g., $\bar C_\mathrm{broken}\approx 0$).

\begin{algorithm}[H]
\DontPrintSemicolon
 Set $\mathbf{\varepsilon}_{0,0}: = \mathbf 0$ everywhere\;
 Solve phase field sub-problem for $\phi_{0,0}$ with $\mathbf{\varepsilon}=\mathbf{\varepsilon}_{0,0}$ and $\phi_\mathrm{con.}(\phi_\mathrm{init.})$\;
 \For{$s \in [1,2,...,s_\mathrm{max.}]$}{
 $\mathbf{\bar \varepsilon}_s := \mathbf{\bar \varepsilon}_{s-1} + \Delta \mathbf{\bar \varepsilon}$\;
 \While{$\Delta \phi > \mathrm{Tol}_\mathrm{AM}$}{
 Solve elasticity sub-problem for $\mathbf{\varepsilon}_{s,n}$ with $\mathbf{\bar \varepsilon}=\mathbf{\bar \varepsilon}_s$ and $\phi=\phi_{s,n}$\;
 Solve phase field sub-problem for $\phi_{s,n+1}$ with $\mathbf{\varepsilon} = \mathbf{\varepsilon}_{s,n}$, $\phi_\mathrm{con.}(\phi_{s,0})$, and $\eta/\Delta t = 0$\;
 $\Delta \phi:= ||\phi_{s,n+1}-\phi_{s,n}||_1$\;
 $n:=n+1$
  }
  Set $\mathbf{\varepsilon}_{s+1,0} := \mathbf{\varepsilon}_{s,n-1}$ and $\phi_{s+1,0} := \phi_{s,n}$ \;
  Calculate $\bar C$ from $\mathbf{\varepsilon}_{s+1,0}$ and $\phi_{s+1,0}$ \;
  \If{$\bar C < C_\mathrm{broken}$}{\rm{Break}}
 }
 \caption{Alternating minimization \label{alg:alternating-minimization}}
\end{algorithm}

\subsubsection{Time-dependent evolution}
The two main differences between the time-dependent evolution (Algorithm \ref{alg:time-discretized-nocontrol} below) and alternating minimization are that the factor $\eta/\Delta t$ is non-zero and $\phi_\mathrm{con.}$ is updated after each pair of sub-problem solves rather than after the convergence of an outer loop.
Our implementation of this method limits the amount of crack growth per step by adapting $\Delta t$ via the inner while-loop in Alg.\ \ref{alg:time-discretized-nocontrol}.
Phase field sub-problem solves are only accepted once the time step $\Delta t$ has been lowered such that $\Delta \phi = ||\phi_{n+1}-\phi_{n}||_1$ is less than an upper bound $(\Delta \phi)_\mathrm{max.}$ or it has reached its own lower bound $(\Delta t)_\mathrm{min.}$.
The time step is allowed to increase again once $\Delta \phi < (\Delta \phi)_\mathrm{max.}/2$ up to a maximum of $(\Delta t)_\mathrm{max.}$, and we do not increment $\mathbf{\bar \varepsilon}$ again until we have both a large time step ($\Delta t_n \ge (\Delta t)_\mathrm{max.}$) and a small change in $\phi$ ($\Delta \phi < (\Delta \phi)_\mathrm{min.}$).
This method is able to accommodate large changes in $\Delta t$ because nothing in our simulations depends on the value of $t$ itself.
By incrementing $\mathbf{\bar \varepsilon}$ independently of the value of $t$, this method avoids a type of strain-rate dependent overstress commonly observed in the literature \cite{miehe_thermodynamically_2010,bilgen_crack-driving_2019}, but it can introduce a `stepping' phenomenon into stress-strain curves when $\phi$ evolves significantly before fracture.
These choices in the design of our time-discretized algorithm are intended to efficiently approach time-continuous fracture and thereby provide a clearer contrast with the near-equilibrium method, in which evolution is also limited by adaptive changes to $\mathbf{\bar \varepsilon}$.

\begin{algorithm}[H]
\DontPrintSemicolon
 Set $n:=0$, $\mathbf{\bar \varepsilon}_{0}: = \mathbf 0$, $\Delta t_0: = (\Delta t)_\mathrm{max.}$, and $\bar C \gg \bar C_\mathrm{broken}$\;
 \While{$\bar C > \bar C_\mathrm{broken}$}{
 Solve elasticity sub-problem for $\mathbf{\varepsilon}_{n}$ with $\mathbf{\bar \varepsilon}=\mathbf{\bar \varepsilon}_n$ and $\phi=\phi_{n}$\;
  \While{\rm{True}}{
  Solve phase field sub-problem for $\phi_{n+1}$ with $\mathbf{\varepsilon} = \mathbf{\varepsilon}_{n}$, $\phi_\mathrm{con.}(\phi_{n})$, and $\Delta t = \Delta t_n$\;
  $\Delta \phi: = ||\phi_{n+1}-\phi_{n}||_1$\;
  \eIf{$\Delta \phi < (\Delta \phi)_\mathrm{max.}$ \rm{or} $\Delta t_n \le (\Delta t)_\mathrm{min.}$}
  {
  \If{$\Delta \phi < (\Delta \phi)_\mathrm{max.}/2$}
   {$\Delta t_n: = 2\Delta t_n$\;}
  \rm{Break}\;}
  {$\Delta t_n: = \Delta t_n/2$}
  }
  \If{$\Delta \phi < (\Delta \phi)_\mathrm{min.}$ \rm{and} $\Delta t_n \ge (\Delta t)_\mathrm{max.}$}{
  {$\mathbf{\bar \varepsilon}_{n+1} := \mathbf{\bar \varepsilon}_{n} + \Delta \mathbf{\bar \varepsilon}$\;}}
  Set $\Delta t_{n+1} :=\Delta t_n$ and calculate $\bar C$ from $\mathbf{\varepsilon}_n$ and $\phi_{n+1}$\;
  $n:=n+1$\;
 }
 \caption{Time-discretized algorithm \label{alg:time-discretized-nocontrol}}
\end{algorithm}

\subsubsection{Near-equilibrium algorithm}
Instead of a path-following algorithm where the entire problem is directly coupled to a path-following constraint, we employ a heuristic algorithm that rescales $\mathbf{\varepsilon}$ via an explicit formula intended to keep the driving force for evolution of $\phi$ in Eq.\ \eqref{eq:time_evolution}, $-F_\phi$, at or below an upper bound $(-F_\phi)_\mathrm{max.}$.
Algorithm \ref{alg:near-equilibrium} describes the overall control flow for our near-equilibrium evolution method, and the rescaling procedure is in lines 4-10.
The rescaling is done based on values of $-F_\phi$ and the crack driving force term $-h'(\phi)\psi^+_0(\mathbf{\varepsilon})$ at the grid point $\mathbf x^*$ where $-F_\phi$ is at a maximum.
Taking advantage of the fact that $\psi^+_0(\mathbf{\varepsilon})$ is degree-two homogeneous in $\mathbf{\varepsilon}$ (i.e., that $\psi^+_0(\gamma \varepsilon) = \gamma^2 \varepsilon$) in this small-strain context, line 8 solves for the scaling factor $\gamma$ that sets $-F_\phi = (-F_\phi)_\mathrm{max.}$ at $\mathbf x^*$ if the entire strain field undergoes the rescaling $\mathbf{\varepsilon}_n = \gamma \mathbf{\varepsilon}_n$ in line 10.
Homogeneity also explains why this rescaling produces valid solutions to Eq.\ \eqref{eq:stationarity_u}: despite being highly non-linear, the expressions for the stresses in Eqs.\ \eqref{eq:stress_spectral} and \eqref{eq:stress_voldev} are still degree-one homogeneous in $\mathbf{\varepsilon}$.
In line 9, the maximum increase in a component of $\mathbf{\bar \varepsilon}$ via rescaling is limited to be less than or equal to the the largest component of the strain increment $\Delta \mathbf{\bar \varepsilon}$.
Lines 4-5  allow the rescaling to be triggered only after the crack driving force term $-h(\phi)\psi_0^+(\mathbf{\varepsilon})$ reaches a threshold value.

\begin{algorithm}[H]
\DontPrintSemicolon
 Set $n:=0$, $\mathbf{\bar \varepsilon}_{0} := \mathbf 0$, $\phi_0 := \phi_\mathrm{init.}$, and $\bar C \gg C_\mathrm{broken}$\;
 \While{$\bar C > C_\mathrm{broken}$}{
 Solve elasticity sub-problem for $\mathbf{\varepsilon}_{n}$ with $\mathbf{\bar \varepsilon}=\mathbf{\bar \varepsilon}_n$ and $\phi=\phi_{n}$\;
  \If{$\max \left[-F_\phi (\phi_n, \mathbf{\varepsilon}_n)\right] > (-F_\phi)_\mathrm{max.}$ \rm{and} $\max \left[-h(\phi_n)\psi_0^+(\mathbf{\varepsilon}_n)\right] > (-h\psi_0^+)_\mathrm{thresh.}$}
  {Flag:=True}
  \If{$\max \left[-F_\phi (\phi_n, \mathbf{\varepsilon}_n)\right] > (-F_\phi)_\mathrm{max.}$ \rm{and Flag is True}}
  {
  Find $\mathbf x^* = \mathrm{arg} \max_{\mathbf{x}} \left[-F_\phi(\mathbf x)\right]$\;
  $\gamma := \sqrt{\frac{(-F_\phi)_\mathrm{max.} - \left[ -F_\phi(\phi_n, \mathbf{\varepsilon}_n) +h'(\phi_n)\psi^+_0(\mathbf{\varepsilon}_n)\right]|^{\mathbf x^*}}{h'(\phi_n)\psi^+_0(\mathbf{\varepsilon}_n) |^{\mathbf x^*}} }$\;
  $\gamma: = \min\left(\gamma, 1+ \frac{\max(|\Delta \mathbf{\bar \varepsilon}|)}{\max (|\mathbf{\bar \varepsilon}_n|)} \right)$\;
  Set $\mathbf{\varepsilon}_n := \gamma \mathbf{\varepsilon}_n$ and $\mathbf{\bar \varepsilon}_n := \gamma \mathbf{\bar \varepsilon}_n$\;
  }
  Solve phase field sub-problem for $\phi_{n+1}$ with $\mathbf{\varepsilon} = \mathbf{\varepsilon}_{n}$, $\phi_\mathrm{con.}(\phi_{n})$, and $\Delta t = (\Delta t)_\mathrm{max.}$\;
  Calculate $\bar C_n$ from $\mathbf \varepsilon_n$ and $\phi_{n+1}$\;
  \If{$||\phi_{n+1}-\phi_{n}||_1 < (\Delta \phi)_\mathrm{min.}$}
  {$\mathbf{\bar \varepsilon}_{n+1}: = \mathbf{\bar \varepsilon}_{n} + \Delta \mathbf{\bar \varepsilon}$\;}
  $n:=n+1$
 }
 
 \caption{Near-equilibrium algorithm \label{alg:near-equilibrium}}
\end{algorithm}

If $\mathbf x^*$ remains the point with the largest value of $-F_\phi$ after rescaling, then Algorithm \ref{alg:near-equilibrium} enforces an upper bound on $-F_\phi$ within the entire system, limiting the extent to which it can be shifted out of equilibrium.
It is not difficult to provide theoretical counterexamples where this bound would be violated, but such behavior was rarely observed during our simulations.
We did observe snap-back events in the stress-strain curve that appeared to be spurious (e.g., during otherwise stable crack growth in a homogeneous domain), but these events temporarily inhibit the evolution of $\phi$ and thus should not affect the crack path.
Another concern is getting stuck in a cycle of loading and unloading with exclusively reversible evolution (e.g., with crack-set irreversibility), but this was not encountered in the simulations shown here.
The relationship between $(-F_\phi)_\mathrm{max.}$ and global measures of evolution such as $||\phi_{n+1}-\phi_{n}||_1$ is variable and dependent on both the choice of model parameters in $F_\phi$ and the grid resolution.
Our approach is in some respects related to the staggered path-following method introduced by Singh et al.\ \cite{singh_fracture-controlled_2016}; we would characterize our approach as being simpler to implement (since the strain is rescaled outside of the sub-problem solvers), but subject to the above drawbacks.
A direct comparison of path-following approaches is outside the scope of this work.

\subsection{Non-Dimensionalization and Simulation Parameters} \label{sec:parameters}
Since this work is focused on comparing methods rather than examining a particular material system, we consider all dimensional quantities in terms of model parameters rather than physical units.
We scale length by the regularization parameter $\ell$.
Per Eqs.\ \eqref{eq:AT1_analytical} and \eqref{eq:AT2_analytical}, $\ell$ is the inverse of the magnitude of the slope of $\phi(x)$ at the crack center, and thus $2\ell$ can be considered an approximate width for the highly damaged `core' of the crack.
There are multiple energy densities that are relevant for scaling, but the most convenient are the fracture energy density $G_c/\ell$ and a reference Young's modulus $E_0=10^4 G_c/\ell$.
The high ratio of $E_0/(G_c/\ell)$ is intended to ensure that fracture occurs at small strains.
In our 2-D systems, integrated energies such as $F_f$ are scaled by $G_c \ell$.
The characteristic time scale for the time-dependent models is $\tilde t = \eta/(G_c/\ell)$.

We scale stresses and strains by the maximum values $\sigma_M$ and $\varepsilon_M$ that they could obtain in a homogeneous material with Young's modulus $E_0$ \cite{pham_gradient_2011}.
In general, these quantities depend on the phase field fracture model (both AT1 vs.\ AT2 and the choice of mechanics model) and the applied loading.
The most relevant case for this work is the AT1 model subject to a strain in which $\varepsilon_{22}$ is positive and the only non-zero component.
In this case, the mechanics models in Eqs.\ \eqref{eq:isotropic_en}-\eqref{eq:vol_dev_en} behave identically, and we have
\begin{equation}
\label{eq:tension_M}
    \psi^+_0 (\varepsilon_M) h'(0) = \frac{3G_c}{8\ell}f'(0)
\end{equation}
\[
   (\lambda + 2\mu) \varepsilon_M^2 = \frac{3G_c}{8\ell}
\]
\begin{equation}
\label{eq:epsilon_M}
    \varepsilon_M = \sqrt{\frac{3G_c}{8\ell(\lambda +2\mu)}}
\end{equation}
\begin{equation}
\label{eq:sigma_M}
    \sigma_M = \sqrt{\frac{3G_c(\lambda +2\mu)}{8\ell}}
\end{equation}
For $\nu=0.2$, we have $(\lambda + 2\mu) = \frac{10}{9}E_0$, which results in $\varepsilon_M = \sqrt{27/(8\times 10^5)}\approx0.005809$ and $\sigma_M = \sqrt{10^5/24}G_c/\ell\approx 64.55G_c/\ell$.

We simulate fracture in 2-D domains of size $L_x=L_y=100\ell$, $L_x=L_y=200\ell$, and $L_x=L_y=400\ell$, with grid sizes that are respectively $N_x=N_y=513$, $N_x=N_y=1023$, and $N_x=N_y=2047$.
These grids result in $\ell/\Delta x \approx 5$, which is comparable to best practice resolutions for finite element discretizations of phase field fracture \cite{bleyer_dynamic_2017}.
For the smaller simulations ($L_x \le 200\ell$), we use relative and absolute tolerances of $10^{-6}$ for both sub-problem solvers and $(\Delta \phi)_\mathrm{min.} = \mathrm{Tol}_\mathrm{AM} = 10^{-3}$ for all three evolution methods.
The time dependent method additionally has $(\Delta t)_\mathrm{max.}=2^{16}\tilde t\approx 6.55\times 10^4\tilde t$, $(\Delta t)_\mathrm{min.}=2^{-16}\tilde t\approx 1.53\times 10^{-5}\tilde t$, and $(\Delta \phi)_\mathrm{max.} = 1.5$, while the near-equilibrium method has $(\Delta t)_\mathrm{max.}=2^{16}\tilde t$, $(-F_\phi)_\mathrm{max.}=0.7G_c/\ell$, and $(-h\psi_0^+)_\mathrm{thresh.}=1G_c/\ell$.
Relaxed tolerances and a larger limiting driving force $(-F_\phi)_\mathrm{max.}$ were used for simulations with $L_x=L_y=400\ell$.
Since we only show crack paths for one set of such simulations, we give these modified conditions alongside the description of the simulations in Section \ref{sec:evolution_random}.
For simulations of tensile fracture, the termination criterion $C_\mathrm{broken}$ has been set to $0.01E_0$.

\subsection{Microstructure Generation} \label{sec:structure_gen}

In this work, we compare the crack paths generated by phase field fracture models in three different types of periodic structure.
The first type of structure consists of domains with uniform material properties into which a crack or defect is incorporated via the initial condition of the phase field.
We consider via this method a periodic version of the standard single-edge notched tension and shear tests \cite{ambati_review_2015,chen_fft_2019} as well as tensile fracture initiated at a small void.
The second and third types of structure employ spatially varying Young's moduli of the form $E(\mathbf x) = E_0 \xi(\mathbf x)$, where $\xi(\mathbf x)$ is constructed from a Gaussian random field to have a mean of approximately unity.
Our second type of structure employs a random field for $\xi$ directly while the third type thresholds $\xi$ into a two-phase structure, sometimes called a ``slit island'' analyis~\cite{mandelbrot_fractal_1984}.
Such two-phase structures have been considered as surrogates for random two-phase systems in materials science, and their geometric characteristics have been extensively studied \cite{Teubner1991random,soyarslan20183d}.

To generate our random structures, we initialize $\xi(\mathbf x)$ at each grid point with values sampled from a Gaussian distribution with zero mean and unit variance.
We then apply a low-pass filter to eliminate Fourier modes with wavelengths smaller than a cutoff wavelength $L_\mathrm{cut}$.
This step determines the distribution of size scales (e.g., interfacial curvatures) present in the microstructure \cite{Teubner1991random}.
To apply the low-pass filter, we take the discrete Fourier transform of $\xi$,
\begin{equation}
    \hat \xi (\mathbf q) = \mathcal F_N \left[ \xi (\mathbf x)\right]
\end{equation}
and set to zero the Fourier components of $\xi$ that have $|q|<2\pi/L_\mathrm{cut}$,
\begin{equation}
    \hat \xi(\mathbf q) := 0,\;\; \forall \mathbf q: |\mathbf q| < \frac{2\pi}{L_\mathrm{cut}},
\end{equation}
The inverse Fourier transform is applied to the filtered $\hat \xi$, resulting in a smoothly varying Gaussian random field with zero mean.
The remaining steps are different between the random field structure and the two-phase structure.

For the random field structure, we multiply the filtered field by a scalar to achieve a specific target standard deviation $\mathrm{STD}_t$ and add one to achieve our target mean,
\begin{equation}
    \xi_r := \xi \frac{\mathrm{STD}_t}{\mathrm{STD}(\xi)} + 1,
\end{equation}
where $\mathrm{STD}(\xi)$ denotes the standard deviation of $\xi$ before rescaling.
Since this procedure does not guarantee that $\xi_r$ is greater than zero, we apply an algebraic sigmoid function to values of $\xi_r$ less than one to smoothly enforce $\xi_r \ge \xi_\mathrm{min.}$,
\begin{equation}
\label{eq:sigmoid}
    \xi_r :=  \frac{|\xi_r-1| |\xi_\mathrm{min.}-1|}{\left(|\xi_r-1|^{10} + |\xi_\mathrm{min.}-1|^{10}\right)} + 1, \;\; \forall \xi_r < 1
\end{equation}
where $\xi_\mathrm{min.}$ is a minimum value for $\xi_r$, taken to be 0.01 here.
This procedure results in a smooth, positive field $\xi_r$ that is primarily characterized by the spectral cutoff wavelength $L_\mathrm{cut}$ and the target standard deviation $\mathrm{STD}_t$.
It is no longer Gaussian-random due to Eq.\ \eqref{eq:sigmoid}, but that fact is of no consequence to the simulations.

For the two-phase structure, we threshold $\xi$ at each point such that
\begin{equation}
    \xi_p := \mathrm{sgn}\;\xi,
\end{equation}
where the sign function $\mathrm{sgn}(\cdot)$ indicates that $\xi_p=1$ for positive $\xi$, $\xi_p=0$ for $\xi=0$, and $\xi_p = -1$ for negative $\xi$.
To avoid ringing artifacts that can occur in spectral discretizations with discontinuous changes in properties between pixels, we smooth the segmented field $\xi_p$ with a finite differences iteration based on the Allen-Cahn equation \cite{allen_microscopic_1979, bueno-orovio_spectral_2006}.
This iteration can be expressed as
\begin{equation}
\label{eq:ACsmooth}
    \xi_\mathbf{k}^{n+1}:= 
    \xi_\mathbf{k}^n - 0.1 \left[ \left( \xi^3 -\xi \right) + \left( 4\xi_{\mathbf k} - \xi_{\mathbf k- (1,0)} - \xi_{\mathbf k- (0,1)} - \xi_{\mathbf k+ (1,0)} - \xi_{\mathbf k+ (0,1)}\right) \right],
\end{equation}
for $\mathbf k \in \mathbb Z_N^2$ ($\xi$ at indices outside of $\mathbb Z_N^2$ is known based on periodicity) and $n=0,1,2,...,20$.
The iteration in Eq.\ \eqref{eq:ACsmooth} results in a structure consisting of two phases with $\xi_p=-1$ and $\xi_p=1$ separated by a diffuse interface approximately $2\Delta x$ wide.
This structure is then scaled to its final values according to
\begin{equation}
    \xi_p := a \xi_p +1,
\end{equation}
where the scalar $a < 1$ determines the ratio $(a+1)/(a-1)$ of the Young's moduli of the bulk phases.

We consider three realizations of the smooth random structure and two-phase structure at size $L_x=L_y=100\ell$ and two realizations of the two-phase structure at size $L_x=L_y=400\ell$.
All of these structures have $L_\mathrm{cut}=6\ell$.
The smooth random structures have a target standard deviation for $\xi$ of $\mathrm{STD}_t=0.3$, while the two-phase structures have $a=0.875$, resulting in a ratio of 15 between the Young's moduli of the phases.
To investigate convergence of the crack path with respect to the ratio $L_\mathrm{cut}/\ell$ in already-generated structures, we upscale $(100\ell)^2$ structures via bivariate cubic spline interpolation.
In this process, we use the SciPy RectBivariateSpline class to interpolate from a $513^2$ grid (the original $N_x=N_y=511$ grid plus extra layers of points to ensure periodicity) to a $1023^2$ or $2047^2$ grid corresponding respectively to a larger $(200\ell)^2$ or $(400\ell)^2$ domain and thus a larger value of $L_\mathrm{cut}$.
The results below will typically present only a single realization for a given condition.
Cases where observations do not generalize to the other realizations are specifically noted.

\section{Results and Discussion}

\subsection{Evolution Method} \label{sec:results_evolution}
In this sub-section, we consider the different evolution methods for phase field fracture and compare how they affect crack paths in elastically heterogeneous materials.
To inform our discussion of the heterogeneous case (and to provide some validation for our numerical methods), we first examine simpler systems that are homogeneous except for a single crack or flaw.
All fracture simulations in this chapter were carried out for the AT1 phase field formulation, damage-type irreversibility, the strain-spectral crack driving force (Eq.\ \eqref{eq:strain_spectral_en}), and the stress-free (isotropic) contact model (Eq.\ \eqref{eq:stress_iso}).

\subsubsection{Homogeneous Material}\label{subsec:homog_ev}
Consider a homogeneous domain of size $L_x=L_y=100\ell$ ($N_x=N_y=511$) with $E = E_0 = 10^4G_c/\ell$ and $\nu = 0.2$.
The crack was introduced into the initial condition of the phase field via a generalization of the analytical solution in Eq.\ \eqref{eq:AT1_analytical},
\begin{equation}
\label{eq:init_crack}
    \phi_\mathrm{init.}(\mathbf x)=  
\begin{cases}
     \left(1-\frac{1}{2\ell}||\mathbf r(\mathbf x)||_2\right)^2,  & \text{if } ||\mathbf r||_2 \leq 2\\
     0, & \text{otherwise}
\end{cases}
\end{equation}
where $\mathbf r = ( \max(x-L_\mathrm{crack}/2,0), y)$ with $L_\mathrm{crack}=50\ell$ the length of the crack.
The small void was introduced using Eq.\ \eqref{eq:init_crack} and $\mathbf r = (x, y)$, making it effectively a crack with zero length.
Uniaxial tensile strain was applied to this domain with increments of
\begin{equation}
\label{eq:loading_tensile}
    \Delta  \mathbf{ \bar \varepsilon} =\left\{\begin{matrix}
0 & 0 \\
0 &  10^{-4}
\end{matrix}\right\}.
\end{equation}

\begin{figure}
	\centering
	\includegraphics[width = 15.5cm]{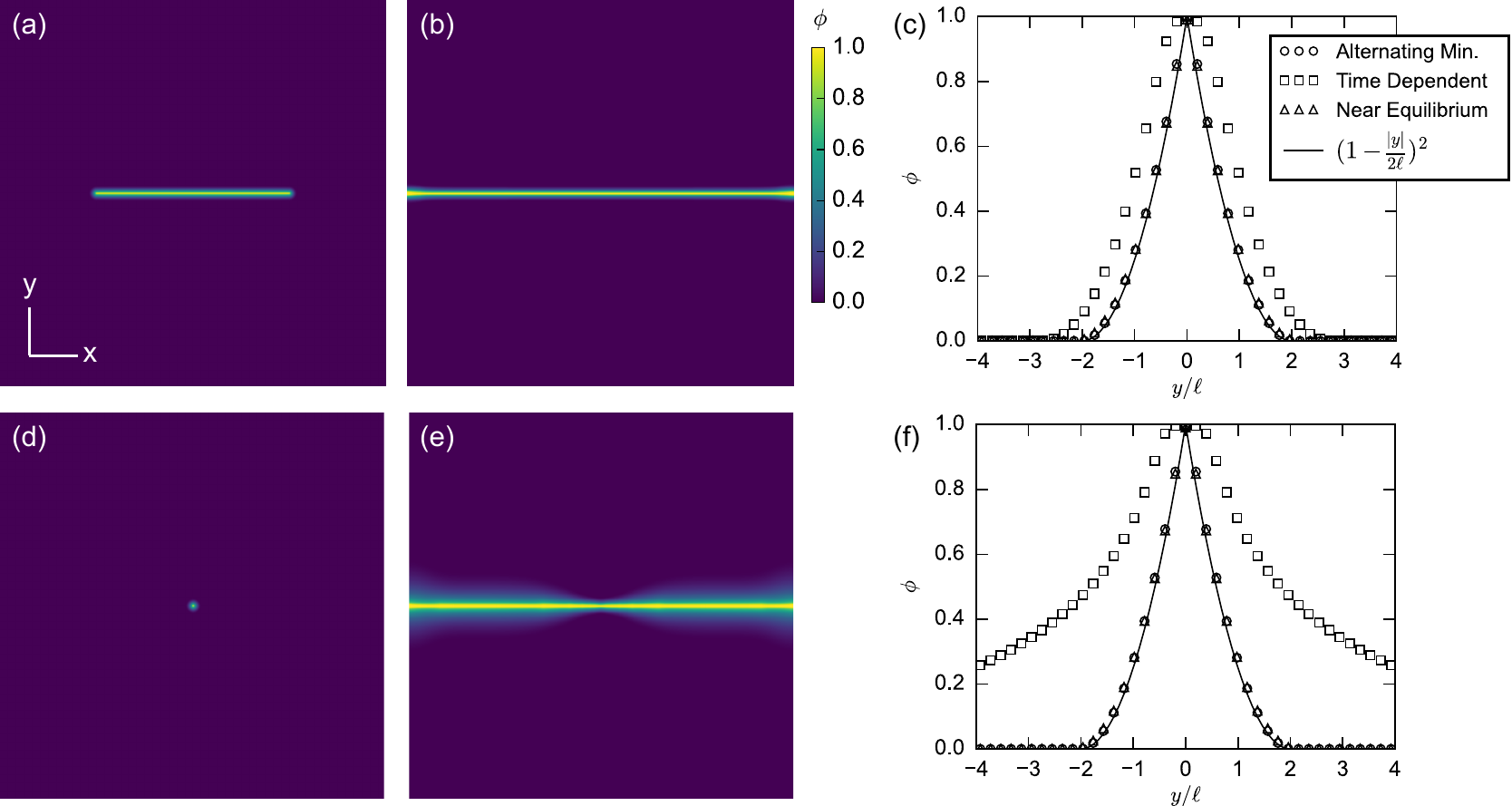}
	\caption{Initial conditions and selected final phase fields for simulations via the three evolution methods (alternating minimization, time dependent, and near equilibrium) in domains with spatially uniform elastic properties and an initial crack (a-c) or small void (d-f) in the phase field.  Pseudocolor plots of the phase field are shown for the initial conditions (a,d) and the final states of the time-dependent simulations (b,e), while the profile of $\phi$ at its final state along the rightmost boundary of the domain (the line $x=50\ell$) is shown for all evolution methods (c,f).  For these profiles, the alternating minimization solution is indicated by circles, the time-dependent solution by squares, and the near-equilibrium solution by triangles. Results for alternating minimization and near equilibrium perfectly coincide in (c,f). The analytical solution is indicated by a solid black line.
}
	\label{fig:homogeneous_tension_phi}
\end{figure}

Figure \ref{fig:homogeneous_tension_phi} compares the initial conditions (ICs) for $\phi$, the final state for the time-dependent evolution method, and the profile of $\phi$ for completed simulations for each evolution method along the vertical line $x=50\ell$ at the right edge of the domain between the crack initial condition (Fig.~\ref{fig:homogeneous_tension_phi}a, b, and c, respectively) and the small void initial condition (Fig.~\ref{fig:homogeneous_tension_phi}d, e, and f, respectively).
The final states for $\phi$ for the alternating minimization and near-equilibrium cases are not shown as they agree with the analytical solution in Eq.\ \eqref{eq:AT1_analytical}: relative errors $\int_\Omega |\phi - \phi_\mathrm{anal.}|\dif\mathbf x/\int_\Omega |\phi_\mathrm{anal.}|\dif\mathbf x$ for the crack and small-void ICs are respectively 5.6\% and 6.6\% for the alternating minimization method and 6.5\% and 6.8\% for the near-equilibrium method.
This agreement with the analytical solution is illustrated qualitatively in Fig.\ \ref{fig:homogeneous_tension_phi}c and f, in which the analytical solution is drawn for comparison.

For the time-dependent method, the final phase field for the crack ICs, shown in Fig.\ \ref{fig:homogeneous_tension_phi}b, also agrees reasonably well with the analytical solution, with a relative error of 10.4\% that is higher than those of the other two methods.
The main difference from the analytical solution in this case is that the crack becomes wider as it nears the domain boundary in Fig.\ \ref{fig:homogeneous_tension_phi}b, resulting in a profile wider than the analytical profile in Fig.\ \ref{fig:homogeneous_tension_phi}c.
A similar behavior is observed with much greater magnitude in the crack for the time-dependent method with the small void ICs, shown in Fig.\ \ref{fig:homogeneous_tension_phi}e.
This crack becomes increasingly broad from the initial void at the center of the domain to the domain boundaries, and the profile of $\phi$ at the domain boundary (Fig.\ \ref{fig:homogeneous_tension_phi}f) is much broader than for any other solution.
The relative difference in $\phi$ between the time-dependent case in Fig.\ \ref{fig:homogeneous_tension_phi}e and the analytical solution is large, at 186\%.

\begin{figure}
	\centering
	\includegraphics[width = 15.5cm]{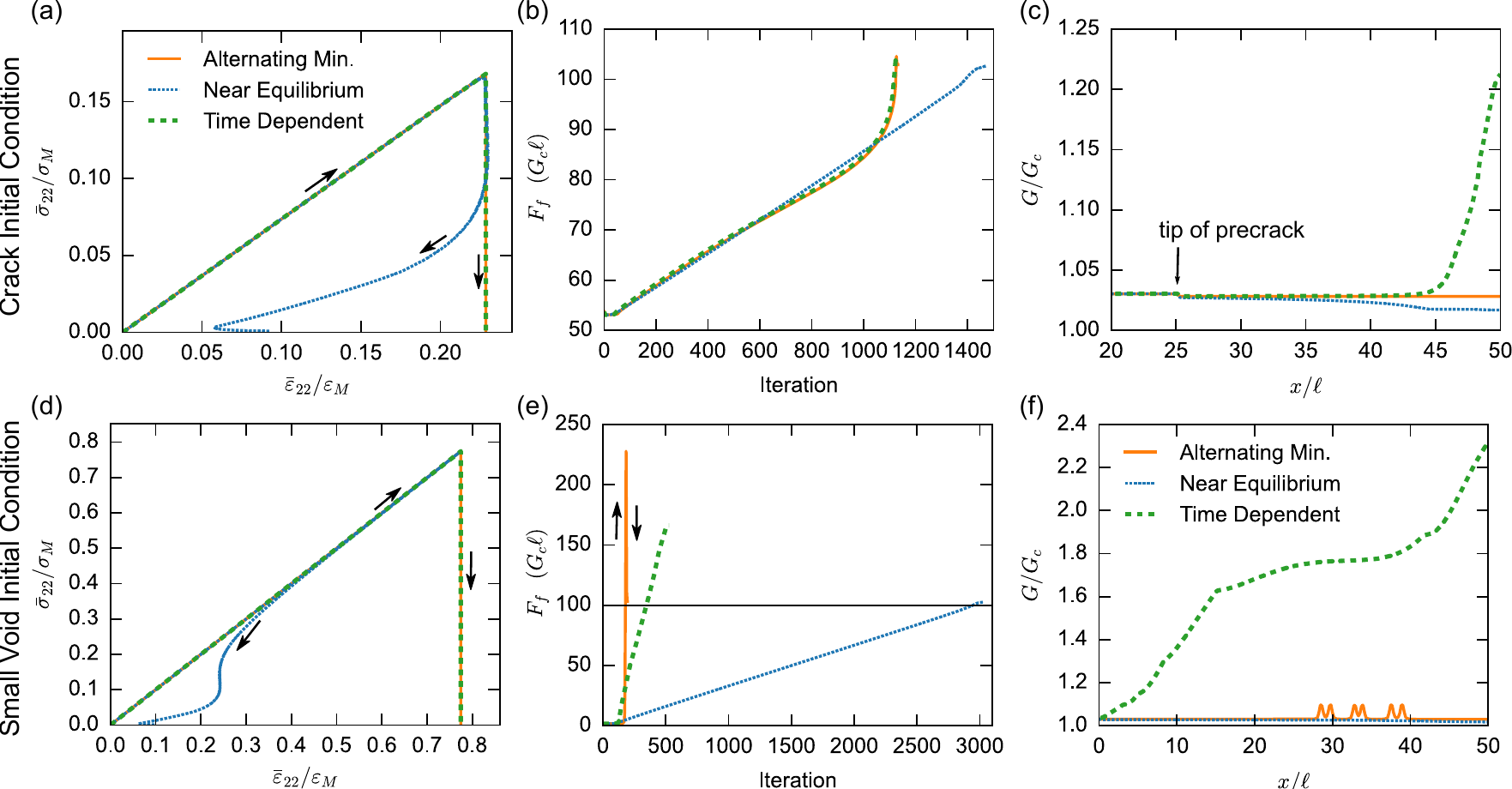}
	\caption{Stress-strain plots (a,d), plots of fracture energy $F_f$ vs.\ iteration (b,e), and plots of energy released per unit length $G$ vs.\ $x/\ell$ for simulations conducted with the three evolution methods (alternating minimization, time-dependent, and near-equilibrium) in domains with spatially uniform elastic properties and an initial crack (a-c) or small void (d-f) in the phase field.  Iteration in (b,e) refers to the time step for the near-equilibrium and time-dependent cases and the inner iteration, indexed cumulatively for all load steps, for the alternating minimization method.  In all plots, the alternating minimization case is indicated by a solid orange line, the time-dependent case by a thick green dashed line, and the near-equilibrium case by a blue finely dashed line.  The analytical final value of $F_f$ in (b,e) is 100, which is indicated in (e) by a thin solid black line.
}
	\label{fig:homogeneous_tension_stats}
\end{figure}

To provide insight into the differences between the time-dependent method and the other methods observed in Fig.\ \ref{fig:homogeneous_tension_phi}, Fig.\ \ref{fig:homogeneous_tension_stats} plots the stress-strain curves, the evolution of the total dissipated fracture energy in the system $F_f$ vs.\ iteration, and the fracture energy per unit length $G/G_c$ vs.\ $x/\ell$ for both the crack and small void ICs.
Average stress $\bar \sigma_{22}$ and strain $\bar \varepsilon_{22}$ are scaled by $\sigma_M$ and $\varepsilon_M$ obtained from Eqs.\ \eqref{eq:epsilon_M} and \eqref{eq:sigma_M}, respectively.
Iteration in Fig.\ \ref{fig:homogeneous_tension_stats}b and e denotes the time step $m$ in Algorithms \ref{alg:time-discretized-nocontrol} and \ref{alg:near-equilibrium} for the time-dependent and near-equilibrium methods respectively, while for the alternating minimization method, Algorithm \ref{alg:alternating-minimization}, it refers to the total number of inner iterations (indexed by $m$) for both the current outer iteration (indexed by $n$) and all previous outer iterations.

The alternating minimization and time-dependent simulations have the same stress-strain curves in Fig.\ \ref{fig:homogeneous_tension_stats}a and d: stress increases linearly with strain until it reaches its maximum value, at which point it decreases to zero at fixed strain.
The slope of the linear regime (i.e., the homogenized elastic constant $\bar C_{2222}=\bar \sigma_{22}/\bar \varepsilon_{22}$ prior to fracture) for the small void IC in Fig.\ \ref{fig:homogeneous_tension_stats}d is nearly $1\sigma_M/\varepsilon_M$, consistent with a nearly homogeneous domain, and the peak stress is relatively high, at $0.77\sigma_M$.
The crack IC results in a less stiff domain, with $\bar C_{2222} =0.65 \sigma_M/\varepsilon_M$, and fractures at a much lower stress of $0.16\sigma_M$.
This difference in fracture stresses is expected, as a crack should induce a singularity in the stress field while a round void should not.

The stress-strain curves for the near-equilibrium case exhibit snap-back, where strain decreases during fracture instead of remaining constant.
We can track the effects of material degradation by noting that the homogenized stiffness $\bar C_{2222}$ at a partially fractured state is the slope of the line between a point on the stress-strain curve and the origin.
For the small void IC case in Fig.\ \ref{fig:homogeneous_tension_stats}d, significant snap-back (almost a 3x reduction in $\bar \varepsilon_{22}$) occurs with very little change in stiffness, whereas stiffness for the crack IC in Fig.\ \ref{fig:homogeneous_tension_stats}a decreases by almost 50\% before significant snap-back occurs.
This behavior could be expected, as the crack that nucleates from the small void during fracture initiation introduces a new stress singularity, greatly reducing the critical value of $\bar \varepsilon_{22}$ needed for crack growth.
Since the time-dependent and alternating minimization simulations experience much higher strains in the small void case than the near-equilibrium simulation when at the same average stiffness, we can say that they undergo crack propagation under overstressed conditions.
Since all of the simulations with the crack IC have no change in strain as stress (and thus stiffness) decreases initially, we can say that they are all similarly close to equilibrium for a large initial part of their evolution.
The near-equilibrium stress-strain curve for the crack IC eventually diverges from the other evolution methods, but the difference in applied strain between them is still small compared to the small void IC case.

To understand how overstress might affect evolution, consider the plots of the fracture energy $F_f$ vs.\ iteration in Figs. \ref{fig:homogeneous_tension_stats}b and e.
For the crack IC case in Fig.\ \ref{fig:homogeneous_tension_stats}b, all three evolution methods give essentially the same amount of crack growth per iteration for the first 600 iterations.
The alternating minimization and time-dependent cases have almost identical evolution thereafter, with the main distinction being a slight drop in $F_f$ at the end of the alternating minimization simulation, which conveniently brings it closer to the ideal value of $100G_c\ell$.
The near-equilibrium case has slower evolution at the end compared to the other two methods, requiring approximately 30\% more iterations to reach its end state.

For the void IC case in Fig.\ \ref{fig:homogeneous_tension_stats}e, evolution of $F_f$ is very different between the three evolution methods.
In the alternating minimization case, $F_f$ peaks at $227.7G_c\ell$ after only 184 iterations before declining rapidly to $103.0G_c\ell$ at 196 iterations.
In the time-dependent case, $F_f$ increases monotonically to $166.2G_c\ell$ over 526 iterations.
In the near-equilibrium case, $F_f$ increases monotonically more slowly than the time-dependent case but stops closer to the ideal value, reaching $102.6 G_c\ell$ after 3024 iterations.
The alternating minimization solution does not appear to suffer any ill effects from its rapid evolution, however, as $G/G_c$ in Fig.\ \ref{fig:homogeneous_tension_stats}c and f remains near unity over the length of the crack, just as it does the near-equilibrium case.
Consistent with the profiles of $\phi(50\ell,y)$ shown in Fig.\ \ref{fig:homogeneous_tension_phi}c and f, $G/G_c$ for the time-dependent simulation only differs from unity near the domain boundary ($x/\ell>45$) for the crack IC in Fig.\ \ref{fig:homogeneous_tension_stats}c, while for the void IC in Fig.\ \ref{fig:homogeneous_tension_stats}f it increases significantly starting from the initial void.

\subsubsection{Randomly Heterogeneous Structures} \label{sec:evolution_random}
\begin{figure}
	\centering
	\includegraphics[width = 15.5cm]{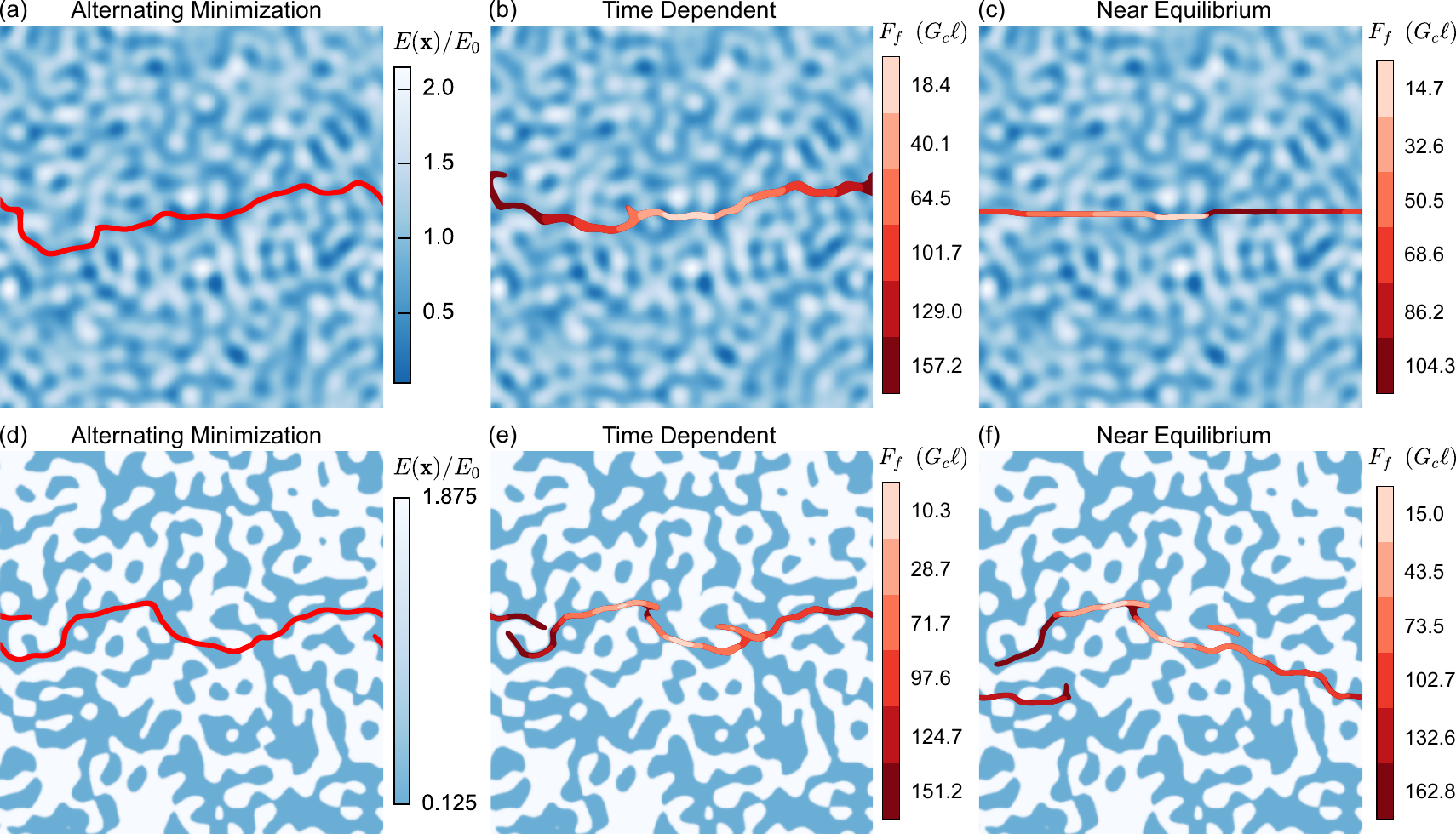}
	\caption{Pseudocolor plots of the scaled Young's modulus $E(\mathbf x)/E_0$ for two randomly heterogeneous structures, one smooth structure (a-c) and one two-phase structure (d-f) overlaid with crack paths with the three evolution methods, alternating minimization (a,d), time-dependent evolution (b,e), and near-equilibrium evolution (c,f).  Both structures have size $L_x=L_y=100\ell$. The crack paths consist of filled contours that depict areas with $\phi\ge0.5$.  The crack paths for the time-dependent and near-equilibrium evolution methods (b, c, e, f) are shaded from light to dark red to show the progression of crack growth, with each level corresponding to the fracture energy $F_f$ depicted in the legend.  The alternating minimization crack paths (a,d) are shown with a solid red color.  Images are shown centered on the crack initiation site (the first point with $\phi>0.95$) for the near-equilibrium evolution method (c,f).
}
	\label{fig:paths_small_evolution}
\end{figure}

Figure \ref{fig:paths_small_evolution} compares crack paths between evolution methods for a smooth random structure and a two-phase structure, both of size $L_x=L_y=100\ell$.
For the smooth random structure, each evolution method produces a qualitatively different crack path.
Both the alternating minimization (Fig.\ \ref{fig:paths_small_evolution}a) and time-dependent (Fig.\ \ref{fig:paths_small_evolution}b) crack paths avoid propagating through regions with high Young's modulus even if it requires them to change direction.
This is in contrast to the near-equilibrium crack path (Fig.\ \ref{fig:paths_small_evolution}), which deviates only slightly from a straight horizontal line.
The time-dependent crack path is notably thicker than both the alternating minimization and near-equilibrium crack paths, which contributes to it having a higher scaled fracture energy $F_f/(G_c \ell)$ at the end of the simulation, with $157.2$ compared to $104.3$ for the near-equilibrium case and $126.5$ for the alternating minimization case.
The time-dependent crack also evolved in both directions simultaneously (as indicated by the intermediate states in Fig.\ \ref{fig:paths_small_evolution}b) while the near-equilibrium crack grew primarily from right to left, eventually re-entering the right side of the periodic domain and continuing to the original initiation site.

For the two-phase structure in Fig.\ \ref{fig:paths_small_evolution}d-f, the crack paths for the different evolution methods are in better qualitative agreement than for the smooth random structure in Fig.\ \ref{fig:paths_small_evolution}a-c.
All of the crack paths in Fig.\ \ref{fig:paths_small_evolution}d-f have evolved significantly in the vertical direction, yielding convoluted crack paths that closely track microstructural features.
In particular, the crack nucleates within regions of low-$E$ phase that separate regions of high-$E$ phase in the vertical direction, and it tends to propagate through the low-$E$ phase where possible.
All of the crack paths agree for approximately half of their extent, deviating eventually because the near-equilibrium crack in Fig.\ \ref{fig:paths_small_evolution}f extends to the lower right while the cracks in Figs.\ \ref{fig:paths_small_evolution}d and e extend to the upper right.
In both the time-dependent (Fig.\ \ref{fig:paths_small_evolution}e) and near-equilibrium (Fig.\ \ref{fig:paths_small_evolution}f) cases, secondary cracks are observed to nucleate and grow, and the old primary crack tip may join with the secondary crack (as in Fig.\ \ref{fig:paths_small_evolution}e and the left side of Fig.\ \ref{fig:paths_small_evolution}f) or bypass it and go in a different direction (as in the right side of Fig.\ \ref{fig:paths_small_evolution}f).
The alternating minimization crack path in Fig.\ \ref{fig:paths_small_evolution}d is missing secondary crack tips that remain in the time-dependent crack path Fig.\ \ref{fig:paths_small_evolution}e; otherwise both evolution methods produce very similar final crack paths.
The final values of $F_f/(G_c \ell)$ are closer for the two-phase structure than for the smooth random structure, with $137.2$ for the alternating minimization method, $151.2$ for the time-dependent method, and $162.7$ for the near-equilibrium method.
(For the time-dependent and near-equilibrium evolution methods, final values for $F_f$ correspond to the largest values in the legends for $F_f$ in Fig.~\ref{fig:paths_small_evolution} and similar figures.)
These differences are due primarily to the different crack paths: the only systematic difference in $F_f$ across all realizations of the two-phase structure is that the time-dependent case has higher $F_f$ than the alternating minimization case.
Differences in crack paths are also more subtle in other realizations compared to Fig.~\ref{fig:paths_small_evolution}d-f.

\begin{figure}
	\centering
	\includegraphics[width = 15.5cm]{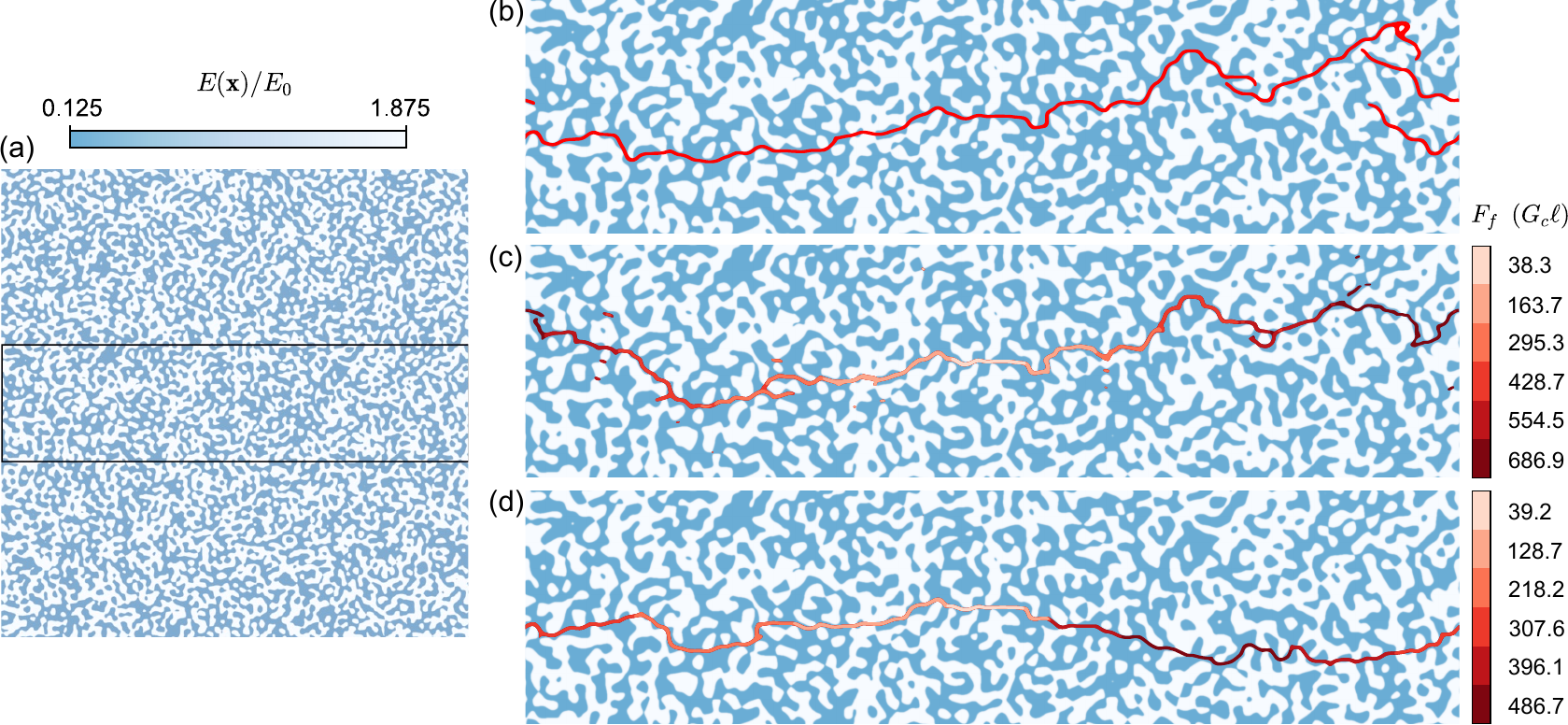}
	\caption{Pseudocolor plots of the scaled Young's modulus $E(\mathbf x)/E_0$ for a two-phase structure with size $L_x=L_y=400\ell$ overlaid with crack paths corresponding to simulations with the three evolution methods, (b) alternating minimization, (c) time-dependent evolution, and (d) near-equilibrium evolution.  The entire structure is plotted in (a) with a black box that indicates the area shown in (b-d).  Centering of the images and depiction of the crack paths is as in Fig.\ \ref{fig:paths_small_evolution}.
}
	\label{fig:paths_large_evolution}
\end{figure}

To further examine possible differences between evolution methods, we also consider fracture of a much larger two-phase structure, with $L_x=L_y=400\ell$ rather than $L_x=L_y=100\ell$.
These simulations use less restrictive convergence tolerances of $10^{-4}$ for the sub-problem solvers, $(\Delta \phi)_\mathrm{min.} = \mathrm{Tol}_\mathrm{AM} = 10^{-2}$ for all evolution methods, and $(-F_\phi)_\mathrm{max.}=1G_c/\ell$ for the near-equilibrium method.
Additionally, the increment of the average strain is smaller, with $\Delta \bar \varepsilon_{22}=2\times 10^{-5}$.
Crack paths for this structure for the three evolution methods are shown in Figure \ref{fig:paths_large_evolution}.
As in the smaller structure in Fig.\ \ref{fig:paths_small_evolution}d-f, there is initially a region of agreement between all three crack paths, but it is much smaller (${\sim20}\%$) relative to the overall crack length.
The alternating minimization and time-dependent crack paths (Fig.\ \ref{fig:paths_large_evolution}b and c, respectively) agree for longer, ${\sim} 40\%$ of their length.
The alternating minimization crack path in Fig.\ \ref{fig:paths_large_evolution}b contains two long secondary cracks that are separated from the longer primary crack.
One secondary crack overlaps with the other cracks over its entire length, while at one location the other secondary crack completely encircles a feature of high-$E$ phase.
For the time-dependent crack path in Fig.\ \ref{fig:paths_large_evolution}c, there many small secondary cracks, particularly once the primary crack has progressed away from its nucleation site.
In contrast, the near-equilibrium crack path in Fig.\ \ref{fig:paths_large_evolution}d has no visible secondary cracks at all, resulting in a lower fracture energy $F_f=486.7G_c\ell$ compared to $643.5G_c\ell$ for the alternating minimization case and $686.9G_c\ell$ for the time-dependent case.

\begin{figure}
	\centering
	\includegraphics[width = 15.5cm]{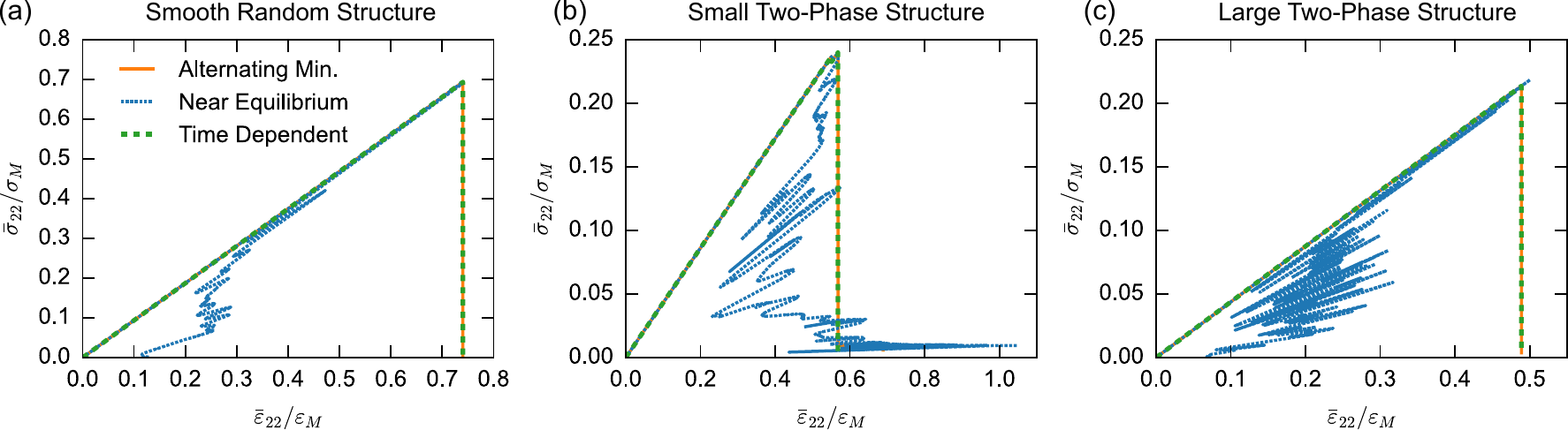}
	\caption{Stress-strain plots comparing the three evolution methods (alternating minimization, time-dependent evolution, and near-equilibrium evolution) for three different structures: (a) the smooth random structure from Fig.\ \ref{fig:paths_small_evolution}a-c, (b) the two-phase random structure from Fig.\ \ref{fig:paths_small_evolution}d-f, and (c) the large two-phase random structure in Fig.\ \ref{fig:paths_large_evolution}.  In all plots, the alternating minimization case is indicated by a solid orange line, the time-dependent case by a thick green dashed line, and the near-equilibrium case by a blue finely dashed line.
}
	\label{fig:stress_strain_evolution}
\end{figure}

Figure \ref{fig:stress_strain_evolution} depicts stress-strain curves corresponding to the crack paths in Figs.\ \ref{fig:paths_small_evolution} and \ref{fig:paths_large_evolution}.
As in Fig.\ \ref{fig:homogeneous_tension_stats}, all three evolution methods share the same linear regime prior to fracture, after which the near-equilibrium method undergoes unloading while the alternating minimization and time-dependent methods evolve the crack with fixed $\bar \varepsilon_{22}$.
The two-phase structures have similar stiffness ($\bar C_{2222}=0.44\sigma_M/\varepsilon_M$ and $\bar C_{2222}=0.42\sigma_M/\varepsilon_M$ for large and small, respectively) and fracture stresses ($0.22\sigma_M$ and $0.24\sigma_M$), while the smooth random structure has significantly higher stiffness $0.94\sigma_M/\varepsilon_M$ and fracture stress $0.69\sigma_M$.
The near-equilibrium stress-strain curves for the large (Fig.\ \ref{fig:stress_strain_evolution}c) and small (Fig.\ \ref{fig:stress_strain_evolution}b) two-phase structures are qualitatively different in that the large structure undergoes more snap-back than the smaller structure.
In this way the large two-phase structure is qualitatively similar to the smooth random structure (Fig.\ \ref{fig:stress_strain_evolution}).
The lesser snap-back in the small two-phase structure can be interpreted as a more rapid degradation of its stiffness.
The large structure might experience relatively less degradation due to the presence of more high-stiffness features along the line of crack growth: the near-equilibrium crack path crosses the high-$E$ phase 14 times in the large structure (Fig.\ \ref{fig:paths_large_evolution}d)and only three times in the small structure (Fig.\ \ref{fig:paths_small_evolution}f).
The small two-phase structure undergoes very high strains and low stresses in Fig.\ \ref{fig:stress_strain_evolution}b at the end of fracture because the tips of the main crack are far apart vertically, and oscillations in the stress-strain curve in this regime are thought to be an artifact of our evolution method.

\subsubsection{Discussion}
Our results indicate that evolution method is an important factor in determining the crack path obtained by phase field simulations of quasi-static brittle fracture in heterogeneous materials.
To understand the origin of the differences between evolution methods, and possible physical interpretations for the different methods, we connect them to our observations for the simple homogeneous examples in Section \ref{subsec:homog_ev}.
Specifically, we note that for the homogeneous structures, all methods resulted in similar cracks when near equilibrium (with the crack IC), but the time-dependent method produced a thicker crack than the others when significant overstresses are present (with the small void IC).
For the heterogeneous structures, all methods behave somewhat similarly near equilibrium (in the small two-phase structure), but the alternating minimization and time-dependent methods behave differently from the near-equilibrium method when significant overstresses are present (away from the crack nucleation site in the smooth random and large two-phase structures).
Recalling the structure of the alternating minimization algorithm and its behavior in Fig.\ \ref{fig:homogeneous_tension_stats}e with the small void IC, its behavior can be explained: as overstress increases, the alternating minimization algorithm evolves $\phi$ increasingly rapidly and non-locally until the two crack tips meet each other (signifying complete fracture), at which point $\phi$ decreases until a local minimizer is obtained.
In our homogeneous examples, only a single local minimizer is available, and the alternating minimization algorithm obtains it successfully.
In the heterogeneous examples, a multiplicity of local minima are available.
Due to its use of a staggered inner iteration, the alternating minimization method selects a crack path (i.e., local minimizer) that may resemble those of the other methods, particularly the time-dependent method.
This is especially true when crack propagation with the alternating minimization method occurs close to equilibrium conditions, such as the homogeneous crack IC example and the small two-phase structure.
When propagation occurs far from equilibrium in a heterogeneous structure, it is not clear that the crack path obtained by the alternating minimization method has a specific physical interpretation.

Given the difficulty of interpreting the minimization approach, the time-dependent evolution is not particularly meaningful as a regularized minimization approach.
Its interpretation as a Ginzburg-Landau-type gradient flow is more useful, primarily because the same interpretation exists for evolution of the phase field in certain models of dynamic fracture \cite{karma_phase-field_2001,hakim_laws_2009}.
The quasi-static time-dependent evolution can be obtained from such dynamic fracture models as the limit of high crack viscosity and/or negligible inertial effects.
Indeed, our simulations with the time-dependent evolution show qualitative features, such as crack widening and branching, that have been observed in phase field simulations of dynamic fracture at high overstress \cite{bleyer_dynamic_2017}.
Crack  branching in dynamic models for phase field fracture is a desirable feature, as it is consistent with experiments.
We conjecture that nucleation of small secondary cracks in the large two-phase structure occurs via a similar mechanism, namely delocalized evolution of the phase field due to overstress (see e.g., Ref.~\cite{scheibert_brittle-quasibrittle_2010} for a possible analog in experiments).
The real questions regarding the time-dependent evolution method are 1) whether the high-viscosity/negligible inertia limit is realistic and 2) whether it is appropriate to label evolution with such a method as `quasi-static'.
Regarding the first question, we note only that the zero-viscosity limit is more commonly considered in recent work on dynamic fracture \cite{bourdin_time-discrete_2011,bleyer_dynamic_2017}.
Regarding the second, the time-dependent method clearly contains physics corresponding to overstress that are absent from the near-equilibrium method but present in dynamic fracture.
Perhaps `quasi-dynamic' fracture would be a more appropriate term for the time-dependent method.

The near-equilibrium evolution appears to be the only method remaining for obtaining accurate crack paths for quasi-static fracture of the types of heterogeneous structure we have considered here.
We do not wish to overstate the applicability of this result.
Deviation from the near-equilibrium crack path appears to depend on overstress and the heterogeneity of the structure, and there may be broad classes of heterogeneous structures that do not induce the differences between evolution methods that we have observed.
Furthermore, minimization methods that seek a global rather than local minimizer for the phase field fracture system \cite{bourdin_numerical_2007,bourdin_variational_2008} provide a fundamentally different piece of information than the quasi-static crack path, which can be interpreted as a lower bound on fracture toughness.
We note that the global minimizer for heterogeneous elasticity with homogeneous local fracture energy is a straight crack.
Our simulations show that none of the evolution methods evolve towards this global minimizer for the two-phase structure; the systems instead appear to naturally evolve to local minimizers with substantially higher dissipated fracture energies than the $L_x G_c\ell$ expected for a straight crack.

\subsection{Mechanics Formulation}
We consider five mechanics models in total: the three variational models introduced in the Background section (isotropic, strain-spectral splitting, and volumetric-deviatoric splitting), plus two non-variational models in which we pair the strain-spectral crack driving force with contact formulations based on the isotropic and volumetric-deviatoric elastic energy densities.
We will refer to these non-variational models by abbreviations of the form `driving force model/contact model', resulting in, respectively, the strain-spectral/stress-free model and the strain-spectral/vol.-dev.~model.
Unless specified otherwise, simulations in this section are conducted with the AT1 model with damage irreversibility evolved by the near-equilibrium method.
Before considering the simulation results themselves, we briefly consider how the mechanics formulations behave analytically when exposed to different strain states.

\subsubsection{Analysis of Mechanics Formulations}
Consider a system in plane stress with principal strains $\mathbf{\varepsilon}^1 = a$ and $\mathbf{\varepsilon}^2 = -a-b$ in the plane, where $a>0$ and $b\ge 0$.
This corresponds to the superposition of a shear strain $a$ and a uniaxial compressive strain $b$.
Since all three variational mechanics formulations have the same degradation function $h(\phi)$, the difference between their crack driving forces $\psi(\phi,\mathbf{\varepsilon})$ lies in their values for the coupled part of the elastic energy, $\psi^+_0$ in Eq.~\eqref{eq:psi_schema}.
These are $\psi^+_0 = \frac{1}{2}\lambda b^2+2\mu(a^2 + ab + b^2/2)$ for the isotropic model, $\psi^+_0 = \mu a^2$ for the strain-spectral split, and $\psi^+_0 = 2\mu(a^2 + ab + 5b^2/18)$ for the volumetric-deviatoric split.
For this loading, one can say generically that $(\psi^+_0)_\mathrm{iso.} \ge (\psi^+_0)_\mathrm{vol.-dev.} > (\psi^+_0)_\mathrm{spectral}$, with equality between the isotropic and volumetric-deviatoric driving forces for pure shear ($b=0$).
The strain-spectral driving force is the only one with no contribution from the compressive strain $b$.
This is desirable from a theoretical perspective, since classical theories for the direction of crack propagation \cite{hutchinson_mixed_1991,hodgdon_derivation_1993} only allow propagation in directions subject to tension.

For undamaged material ($\phi=0$), all three models return the same stresses for a given strain.
We therefore compare the models at a point where $\phi=1$, where stresses  contain only their $\partial \psi_0^-/\partial \mathbf{\varepsilon}$ term.
Such a point corresponds to the center of a crack, and the stresses there correspond to a model for contact of the crack faces \cite{amor_regularized_2009,freddi_regularized_2010}.
These contact models are limited because they lack explicit information about the crack's direction or its surface normal vector, but we can still assess their effects in the context of contact by aligning the system coordinates to the normal vector of the crack surfaces.

Consider a straight crack normal to the $y$-axis.
Instead of simulating the entire domain for this scenario (see, e.g., Ref.~\cite{zhang_assessment_2022} for this case), we consider analytically the response of a point with $\phi=1$ to an imposed local strain.
Given a compressive strain along the $y$-axis, one would expect a contact model to yield a compressive stress.
This is the case for the strain-spectral and volumetric-deviatoric splits, but not for the isotropic model, which is stress free,
\[
    \mathbf{\varepsilon}=\left\{\begin{matrix}
0 & 0 \\
0 &  -b
\end{matrix} \right\}
\]
\begin{equation}
 \mathbf{\sigma}_\mathrm{iso.}=\mathbf 0,\;\;
 \mathbf{\sigma}_\mathrm{spectral}=\left\{\begin{matrix}
-\lambda b & 0 \\
0 &   -(\lambda+2\mu) b
\end{matrix} \right\},\;\;
\mathbf{\sigma}_\mathrm{vol.-dev.}= (\lambda + 2\mu/3) \left\{\begin{matrix}
-b & 0 \\
0 &  -b
\end{matrix} \right\},
\end{equation}
The strain-spectral model is also the only one where the stress response matches the undamaged material since the volumetric-deviatoric model results in lower $\sigma_{yy}$ stress.

For a mixed strain state with tensile strain along the $y$-axis, one would expect zero stress because the tensile strain would bring the crack faces out of contact.
This is the case for the isotropic model and the volumetric-deviatoric split,
\[
    \mathbf{\varepsilon}=\left\{\begin{matrix}
0 & b \\
b & 2b
\end{matrix} \right\}
\]
\begin{equation}
 \mathbf{\sigma}_\mathrm{iso.}=\mathbf 0,\;\;
 \mathbf{\sigma}_\mathrm{spectral}= \frac{\mu}{\sqrt{2}}\left\{\begin{matrix}
-b & (\sqrt{2}-1)b \\
(\sqrt{2}-1)b &    (2\sqrt{2}-3)b
\end{matrix} \right\},\;\;
\mathbf{\sigma}_\mathrm{vol.-dev.}= \mathbf 0.
\end{equation}
The strain-spectral split, on the other hand, retains a significant amount of positive shear stress and introduces new compressive axial stresses in both the $x$- and $y$-directions.
Since the strain-spectral split does not remove shear stresses regardless of the presence of tensile strains, we consider it a `fixed' contact, in contrast to the frictionless behavior of volumetric-deviatoric split \cite{amor_regularized_2009} and the stress-free crack simulated by the isotropic model.
While it does not model contact in compression, the stress-free crack is in fact a common assumption for stress analysis of cracks \cite{zehnder_fracture_2012,rice_mathematical_1968} and sharp-crack models for fracture \cite{larralde_shape_1995,ramanathan_quasistatic_1997,katzav_fracture_2007,lebihain_effective_2020}.
Our analytical observations here are consistent with recent simulations of simple compression and shear by Zhang et al.~\cite{zhang_assessment_2022}.

\subsubsection{Mode II Fracture of a Homogeneous Material} \label{sec:modeII_homogeneous}
Comparisons of mechanics models in literature often examine fracture of a pre-cracked specimen with uniform properties under in-plane shear (mode II) loading \cite{ambati_review_2015, bilgen_crack-driving_2019,zhang_assessment_2022}.
Conditions for these simulations are difficult to replicate directly with periodic boundary conditions.
Fortunately, loading via an applied average shear strain is similar to a classic experiment by Erdogan and Sih \cite{erdogan_crack_1963}, where a distributed shear was applied to a cracked PMMA plate away from the crack (Fig.\ 9 ibid.).
To match this experiment, we simulate fracture within a domain with $L_x = L_y=200\ell$ containing a horizontal crack of length $20\ell$ imposed in either the phase field or the Young's modulus $E(\mathbf x)$.
The crack in $E(\mathbf x)$ is obtained by taking $E(\mathbf x) = h(\phi_\mathrm{init.})E_0$ with $\phi_\mathrm{init.}$ from Eq.\ \eqref{eq:init_crack}, while the phase field crack uses $\phi_\mathrm{init.}$ as the initial condition directly.
Apart from the crack, elastic properties are uniform with $E=E_0$ and $\nu=0.4$, which is more representative of PMMA than $\nu=0.2$.

The domain is then strained in pure shear in increments of
\[
\Delta \mathbf{\bar{\varepsilon}} = \left\{\begin{matrix}
0 & 5\times10^{-5} \\
5\times 10^{-5} &  0
\end{matrix}\right\}.
\]
This strain state implies different values of $\sigma_M$ and $\varepsilon_M$ for the stress and strain for fracture of a homogeneous material than were computed in Eqs.\ \eqref{eq:tension_M}-\eqref{eq:sigma_M} for a pure tensile strain.
For pure shear with the strain-spectral split for the elastic energy density, we have
\begin{align}
    2 \mu \bar \varepsilon_{12,M}^2 &= \frac{3G_c}{8\ell},
    \label{eq:shear_epsilon_M} \\
   \bar \varepsilon_{12,M} &= \frac{1}{4}\sqrt{\frac{3G_c}{\ell \mu}}, \\
   \bar\sigma_{12,M} &= \frac{1}{2}\sqrt{\frac{3G_c\mu}{\ell}},
\end{align}
which for $\nu=0.4$ and $E=10^4G_c$ yields $\bar \varepsilon_{12,M}=0.007246$ and $\bar \sigma_{12,M}=51.75G_c/\ell$.

\begin{figure}
	\centering
	\includegraphics[width = 15.5cm]{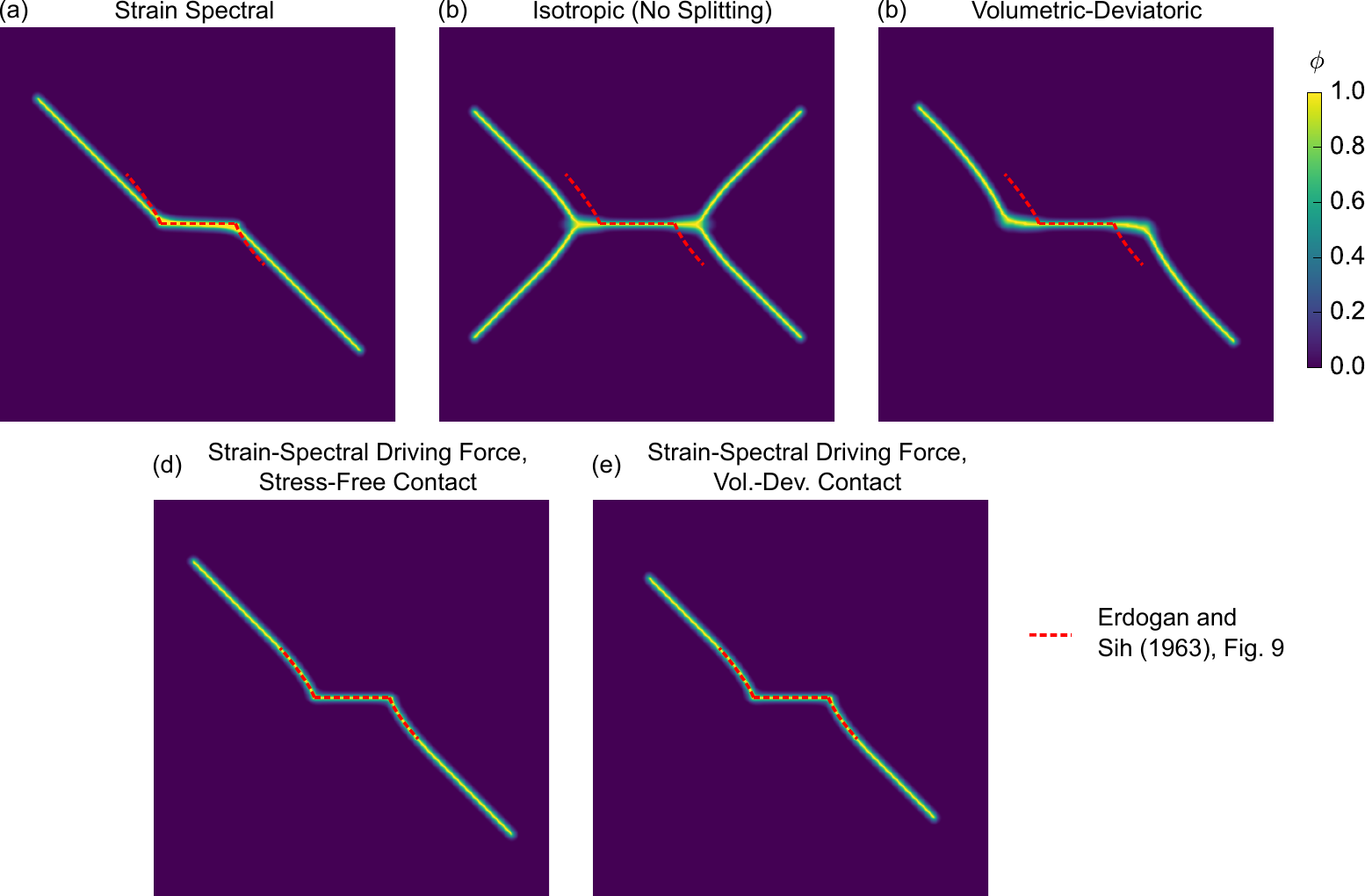}
	\caption{Pseudocolor plots of the phase field resulting from shearing of a homogeneous material with an initial crack in the phase field according to five different mechanics formulations: (a) the strain-spectral variational model, (b) the isotropic variational model, (c) volumetric-deviatoric variational model, (d) the non-variational strain-spectral/stress-free model, and (e) the non-variational strain-spectral/vol.-dev.~model.  Images depict the central area  $[-100\ell,100\ell]^2$, one quarter of the simulation domain.  The red dashed lines indicate a trace of the cracks (one pre-crack and two mode II cracks) in Fig.\ 9 of Erdogan and Sih \cite{erdogan_crack_1963} that has been rotated and rescaled while preserving its aspect ratio.  All sub-figures use the same scaling for this trace.
}
	\label{fig:shear_pf_compare}
\end{figure}

Figure \ref{fig:shear_pf_compare} compares the crack paths resulting from the phase field initial crack to a trace of the three cracks (one pre-crack and two mode II cracks) present in Fig.\ 9 of Erdogan and Sih \cite{erdogan_crack_1963}.
Interestingly, none of the variational models (Fig.\ \ref{fig:shear_pf_compare}a-c) matches the experimental crack path, but the two non-variational models  (Fig.\ \ref{fig:shear_pf_compare}d and e) both fit it very well.
The strain-spectral variational model (Fig.\ \ref{fig:shear_pf_compare}a) results in cracks at a $45^\circ$ angle relative to the pre-crack, which disagrees with the experimental crack path and the angle of $70^\circ$ predicted in Ref.~\cite{erdogan_crack_1963}.
This result does however match previous shear fracture simulations with periodic boundary conditions \cite{chen_fft_2019} and one set of FEM-based simulations \cite{bilgen_crack-driving_2019}.
The isotropic model (Fig.\ \ref{fig:shear_pf_compare}b) results in growth of the pre-crack followed by nucleation of two crack branches per initial crack tip (i.e., four crack branches in total).
The nucleation of these spurious crack branches is expected behavior for the isotropic model \cite{miehe_thermodynamically_2010, ambati_review_2015, bilgen_crack-driving_2019}.
The volumetric-deviatoric model (Fig.\ \ref{fig:shear_pf_compare}c) results initially in growth in the same direction as the pre-crack, but the crack paths eventually change direction and take a path that resembles a scaled-up version of the experimental crack path.

With both the strain-spectral/stress-free and strain-spectral/vol.-dev.\ non-variational models (Fig.\ \ref{fig:shear_pf_compare}d and e, respectively), the mode II cracks propagate directly from the pre-crack with the same angle, overall trajectory, and scale relative to the initial crack as the experimental crack path.
These models also fractured at similar stresses $\bar \sigma_{12}$ of $0.36\sigma_{12,M}$ for the strain-spectral/stress-free model and $0.37\sigma_{12,M}$ for the strain-spectral/vol.-dev.\ model.
This is compared to a much higher fracture stress of $0.56\sigma_{12,M}$ for the strain-spectral variational model and lower fracture stresses of $0.31\sigma_{12,M}$ and $0.33\sigma_{12,M}$ for the volumetric-deviatoric and isotropic variational models, respectively.

\begin{figure}
	\centering
	\includegraphics[width = 15.5cm]{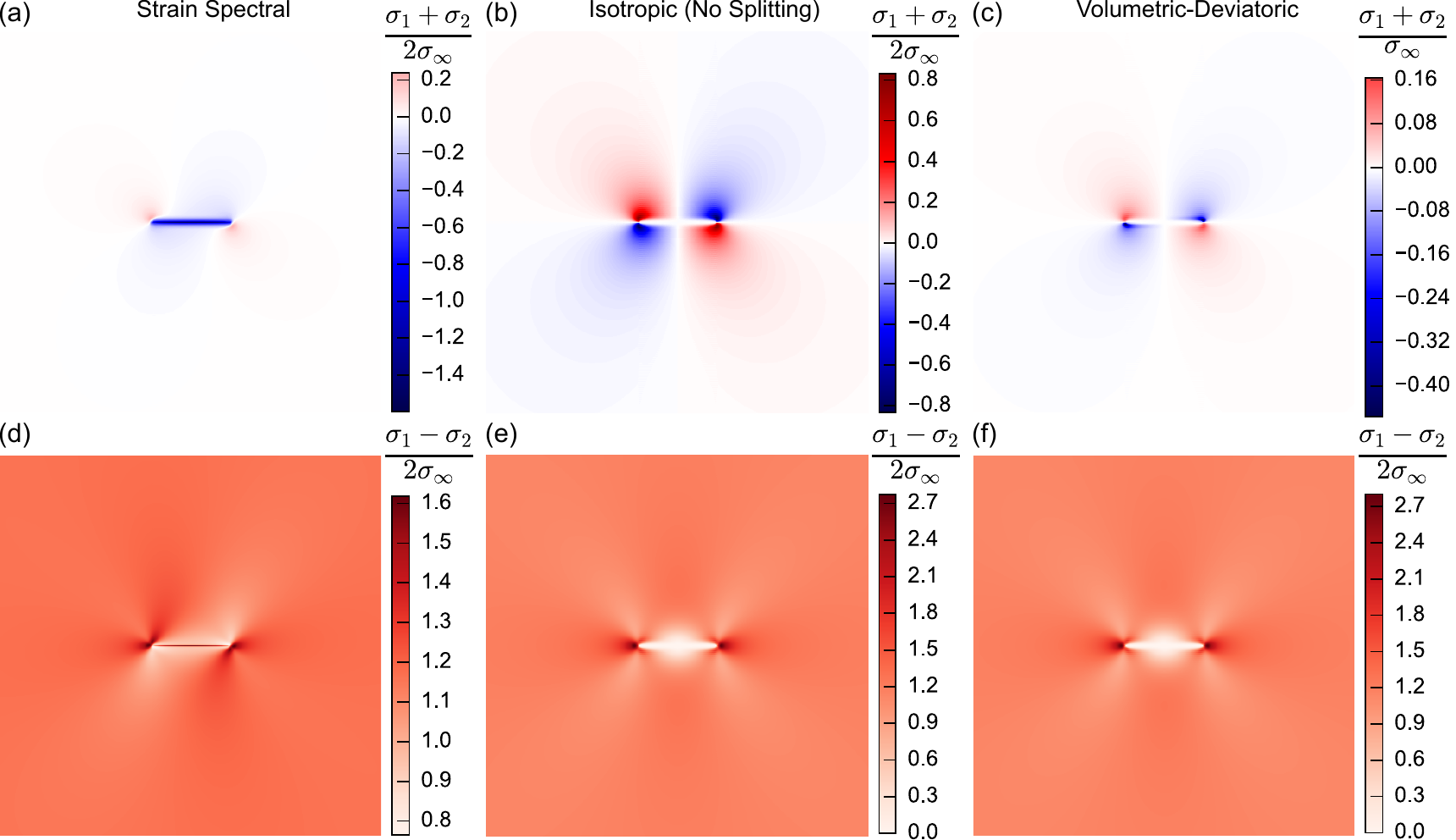}
	\caption{Pseudocolor plots of the trace of the stress $\sigma_1+\sigma_2$ (a-c) and the difference between principal stresses $\sigma_1-\sigma_2$ (d-f) for a phase field crack subjected to an imposed average shear with three mechanics models: (a,d) strain-spectral split, (b,e) isotropic (no split), and (c,f) volumetric-deviatoric split.  The scaling factor $2\sigma_\infty=4\mu \bar\varepsilon_{12}$ corresponds to the far-field value of $\sigma_1+\sigma_2$ induced by the applied average strain in a homogeneous material.
}
	\label{fig:shear_stress}
\end{figure}

To provide insight into the differences in crack path and fracture stress between the simulations in Fig.\ \ref{fig:shear_pf_compare}, Fig.\ \ref{fig:shear_stress} presents pseudocolor plots of the stresses induced by the three contact models prior to significant evolution of the phase field.
In particular, Fig.\ \ref{fig:shear_stress}a-c plots the sum of the principal stresses $\sigma_1$ and $\sigma_2$ (i.e., the trace of the stress tensor), while the difference $\sigma_1-\sigma_2$ is plotted in Fig.\ \ref{fig:shear_stress}d-f.
Both quantities are scaled by the expected far-field value for $\sigma_1-\sigma_2$ based on the applied strain,  $2\sigma_\infty=4\mu\bar \varepsilon_{12}$, which is equal to $14.29G_c/\ell$ in this case.
The stress distribution for the strain-spectral model (Fig.\ \ref{fig:shear_stress}a and c) matches the scenario for a point with $\phi=1$ outlined in the previous sub-section: the applied shear strain induces significant stresses within the crack that correspond to a mixed state of shear and compression, with large negative $\sigma_1+\sigma_2$ and positive $\sigma_1-\sigma_2$.
Net tensile stresses ($\sigma_1+\sigma_2>0$) are concentrated at the crack tips, but the distribution is qualitatively different from the isotropic model and volumetric-deviatoric split in Fig.\ \ref{fig:shear_stress}b and e and \ref{fig:shear_stress}c and f, respectively.
These contact models result in stress-free crack centers and stress concentrations exclusively at the crack tips.
The isotropic model results in a distribution of $\sigma_1+\sigma_2$ that is anti-symmetric about both the $x$- and $y$-axes.
The volumetric-deviatoric split results in a qualitatively similar stress distribution compared to the isotropic model, particularly for $\sigma_1-\sigma_2$, but it has significantly smaller positive (tensile) peak values for the trace $\sigma_1+\sigma_2$.
This difference in stress distribution between the volumetric-deviatoric and isotropic models seems to have had only a minor effect on fracture stress, however, and no noticeable effect on crack path.

\begin{figure}
	\centering
	\includegraphics[width = 15.5cm]{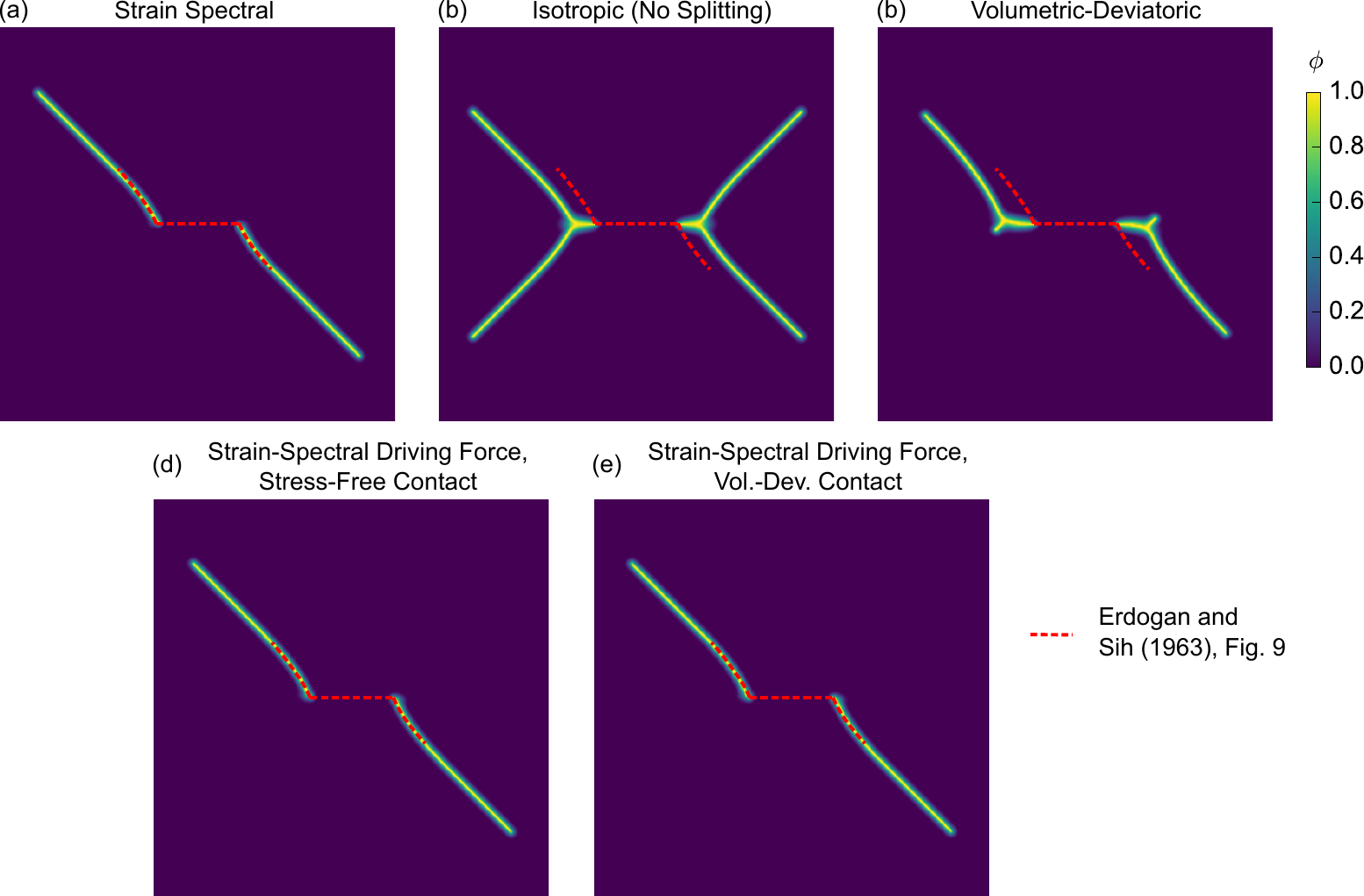}
	\caption{Pseudocolor plots of the phase field resulting from shearing of a homogeneous material with an initial crack in $E(\mathbf x)$ according to five different mechanics formulations: (a) the strain-spectral variational model, (b) the isotropic variational model, (c) volumetric-deviatoric variational model, (d) the non-variational strain-spectral/stress-free model, and (e) the non-variational strain-spectral/vol.-dev.~model.  Images depict the same sub-domain as in Fig.\ \ref{fig:shear_pf_compare}.  The experimental crack path in this figure (red dashed line) has been uniformly rescaled to be 11\% larger than that in Fig.\ \ref{fig:shear_pf_compare}.
}
	\label{fig:shear_cx_compare}
\end{figure}

Simulations with the initial crack in $E(\mathbf x)$ rather than $\phi(\mathbf x)$ serve to further refine our distinction between effects of contact model and crack driving force.
In this case, the contact model contained in the phase field model will affect the mode II cracks that emerge during fracture, but the initial crack in $E(\mathbf x)$ is always stress-free.
The results of these simulations in Fig.\ \ref{fig:shear_cx_compare} are qualitatively similar to those in Fig.\ \ref{fig:shear_pf_compare} with one major exception: the strain-spectral variational model in Fig.\ \ref{fig:shear_cx_compare}a now matches the experimental crack path with the same fidelity as the two non-variational models in Fig.\ \ref{fig:shear_cx_compare}d and e.
All three of these models now have very similar fracture stresses, at $0.47\sigma_{12,M}$ for the variational strain-spectral model and $0.46\sigma_{12,M}$ for both non-variational models.
The large difference in fracture stress between initial conditions in the non-variational cases may be due to the need for nucleation of the new crack when the initial crack is in $E(\mathbf x)$.
Nucleation may also be responsible for more subtle differences between Fig.\ \ref{fig:shear_pf_compare} and Fig.\ \ref{fig:shear_cx_compare}: the scaling factor for the trace of the experimental path is 11\% larger in the latter, and the fits between simulated and experimental crack paths are slightly worse in Fig.\ \ref{fig:shear_cx_compare}a, d, and e than in Fig.\ \ref{fig:shear_pf_compare}d and e.
Fracture stresses for the isotropic and volumetric-deviatoric variational models were both $0.34\sigma_{12,M}$, consistent with the similarity between their crack driving forces under shear that was noted in the previous sub-section.
Note that $\sigma_1-\sigma_2$, which corresponds to shear stress, is maximized in Fig.\ \ref{fig:shear_stress}e and f along the same axis as the initial crack, and it is at this location that crack grows initially in the isotropic and volumetric-deviatoric models in Figs.~\ref{fig:shear_pf_compare} and \ref{fig:shear_cx_compare}.

\subsubsection{Mixed-Loading Fracture of a Randomly Heterogeneous Material}
We now consider how different mechanics models affect crack paths in heterogeneous structures.
Our main interest in this study is tensile fracture, but the differences between mechanics models are greatest for compressive stress states \cite{de_lorenzis_nucleation_2021}.
As a compromise, we consider a mixed loading state with an applied strain increment of
\[
\Delta \mathbf{\bar \varepsilon} = \left\{\begin{matrix}
-5\times10^{-5} & 0 \\
0 &  10^{-4}
\end{matrix}\right\}.
\]
(Poisson's ratio is set to 0.2, as in all other simulations except those of Section \ref{sec:modeII_homogeneous}.)

\begin{figure}
	\centering
	\includegraphics[width = 15.5cm]{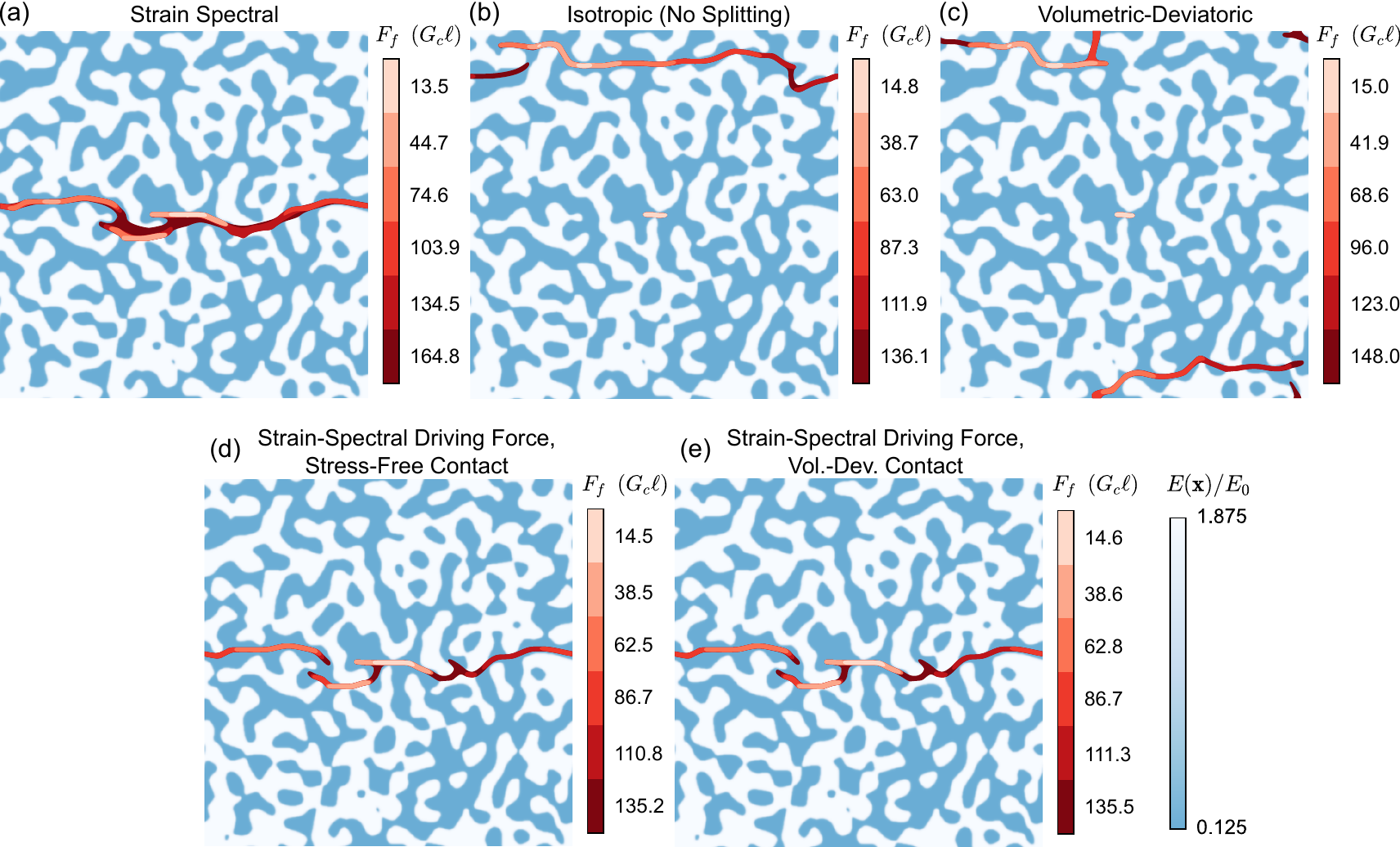}
	\caption{Crack paths resulting from mixed tensile-compressive loading of a two-phase random structure of size $L_x=L_y=100\ell$ via phase field fracture with five different mechanics formulations:  (a) the variational model for the strain-spectral split, (b) the variational model for the isotropic formulation (no split), (c) the variational model for the volumetric-deviatoric split, (d) the strain-spectral/stress-free non-variational model, and (e) the strain-spectral/vol.-dev.~non-variational model.
}
	\label{fig:paths_mixed_elasticity}
\end{figure}

Figure \ref{fig:paths_mixed_elasticity} shows crack paths for the different mechanics models for a two-phase random structure of size $L_x=L_y=100\ell$.
Crack paths for the non-variational models with the strain-spectral driving force in Fig.~\ref{fig:paths_mixed_elasticity}d and e are essentially identical.
The variational strain-spectral crack in Fig.~\ref{fig:paths_mixed_elasticity}a) follows the same path as the non-variational models for much of its evolution, but the crack grows wider in certain locations as the simulation progresses.
This widening results in a substantially higher final fracture energy $F_f$ compared to the non-variational models.
This widening also smooths the crack path, resulting in fewer high-curvature features (kinks or corners) compared to the non-variational models.
The isotropic and volumetric-deviatoric variational models (Fig.~\ref{fig:paths_mixed_elasticity}b and c, respectively) initially nucleate a crack at the same location as the models with the strain-spectral driving force, but growth of this crack is arrested and the domain is eventually perforated by crack growth from other nucleation sites.
Both cases have the same secondary nucleation sites in the upper left of the domain, but their crack paths bifurcate due to crack growth from an additional out-of-plane nucleation site in the volumetric-deviatoric case (Fig.~\ref{fig:paths_mixed_elasticity}c).
Crack growth from secondary nuclei also occurs in the cases with strain-spectral driving force, but close enough to the primary crack that they are able to coalesce.
Figure \ref{fig:paths_mixed_elasticity} represents the greatest contrast in crack paths between different crack driving forces out of the realizations of the two-phase structure that we have simulated; we typically observed smaller differences in other realizations.
However, similarity between the two non-variational models and the crack widening phenomenon for the variational strain-spectral model were observed consistently across realizations, as well as in simulations with purely tensile average strains.

\begin{figure}
	\centering
	\includegraphics[width = 15.0cm]{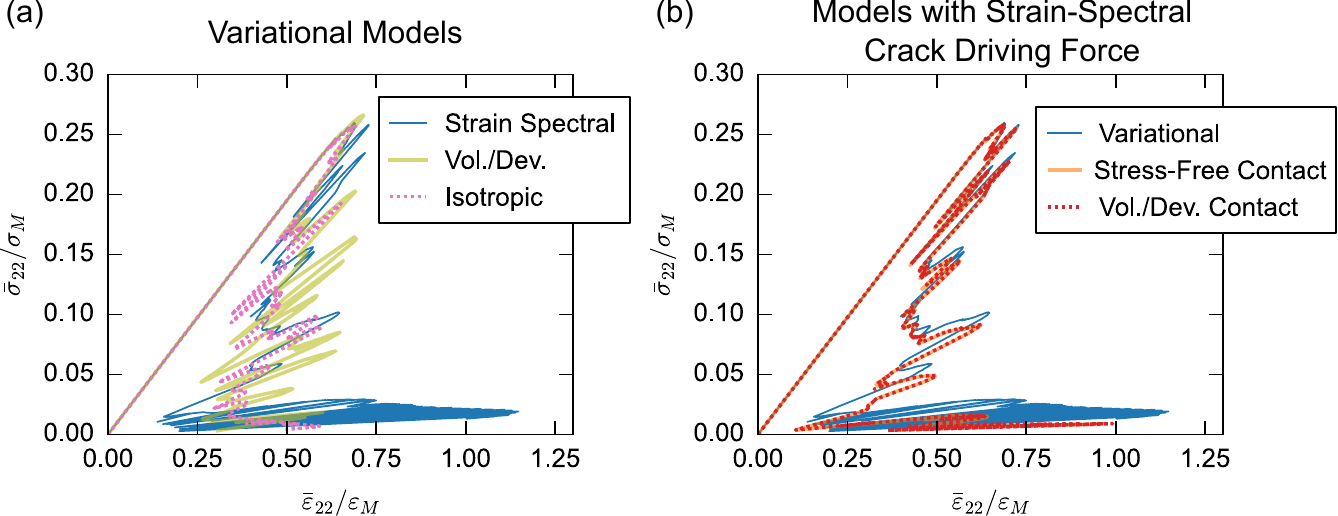}
	\caption{Stress-strain plots corresponding to fracture simulations of the random two-phase structure shown in Fig.\ \ref{fig:paths_mixed_elasticity} with (a) variational mechanics models and (b) mechanics models with the strain-spectral crack driving force.  The curve for the strain-spectral variational model is shown as a thin solid blue line in both plots. (a) also depicts curves for the volumetric-deviatoric split (solid olive line) and isotropic model (pink dashed line). (b) also depicts curves for the strain-spectral/stress-free model (solid orange line) and the strain-spectral/vol.-dev.~model (red dashed line).
}
	\label{fig:stress_strain_mechanics}
\end{figure}

Figure \ref{fig:stress_strain_mechanics} depicts the stress-strain curves for the simulations whose crack paths are plotted in Fig.\ \ref{fig:paths_mixed_elasticity}.
All stress-strain curves show a linear elastic regime followed by a jagged pattern of snap-back events and reloading that is typical for the near-equilibrium evolution method in a heterogeneous structure.
In Fig.\ \ref{fig:stress_strain_mechanics}a, which compares the variational mechanics models, the isotropic and volumetric-deviatoric models result in qualitatively similar stress-strain curves that differ significantly from the strain-spectral curve after the initial snap-back event and from each other after ${\sim}3$ additional snap-back events.
This is an expected consequence of the difference in crack paths in Fig.\ \ref{fig:paths_mixed_elasticity}.
In Fig.\ \ref{fig:stress_strain_mechanics}b, all of the models with the strain-spectral crack driving force have the same pattern of snap-back events for much of their evolution.
The stress-strain curves for the two non-variational models are essentially identical, but they differ from the variational strain-spectral model by having consistently lower stresses, a difference that increases as fracture progresses.
The end of the stress-strain curve for the strain-spectral variational model is characterized by oscillations between low and high strain, and higher strains are needed for fracture compared to the other models.
This oscillatory behavior, not present with the other models, is likely an artifact of our near-equilibrium algorithm and not representative of the equilibrium path.

All of the mechanics models have similar peak $\bar \sigma_{yy}$ stresses, with $0.27\sigma_M$ for the volumetric-deviatoric variational model and $0.26\sigma_M$ for the other models.
This is contrary to the behavior of these models in a homogeneous material, where the compressive strain component would contribute to the crack driving force in the isotropic and volumetric-deviatoric models but not the strain-spectral model.
Following the methodology in Eq.~\eqref{eq:sigma_M}, the strain-spectral crack driving force would have a fracture stress in a homogeneous material that is 12\% higher than the isotropic and volumetric-deviatoric models.

\begin{figure}
	\centering
	\includegraphics[width = 10.0cm]{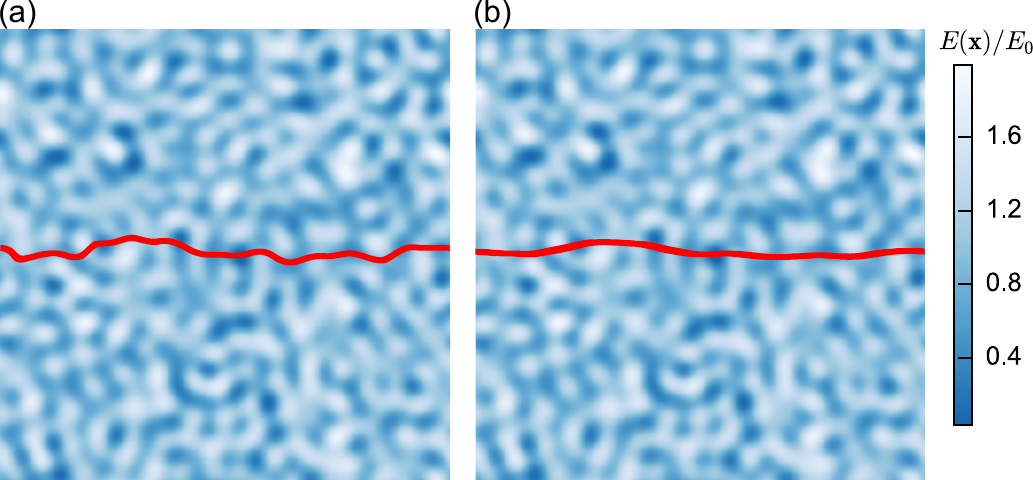}
	\caption{Crack paths (filled $\phi=0.5$ contours) for simulations of fracture of a smooth random structure via the alternating minimization method with (a) the strain-spectral/stress-free mechanics model and (b) the strain-spectral variational model.
}
	\label{fig:paths_alternating_grf}
\end{figure}

Since the crack widening/smoothing phenomenon observed for the strain-spectral variational model in Fig.\ \ref{fig:paths_mixed_elasticity}a is not observed in any other model, we have conducted simulations with additional formulations and evolution methods to test its generality.
Similar widening/smoothing to Fig.\ \ref{fig:paths_mixed_elasticity}a is observed for crack-set irreversibility, the AT2 model, and the time-dependent evolution method.
When alternating minimization is applied to smooth random structures with the strain-spectral variational model, we find minimal widening but substantial smoothing compared to the strain-spectral/stress-free case.
An example of this effect is shown in Fig.~\ref{fig:paths_alternating_grf}.

\subsubsection{Discussion}
Our results suggest that the distinction between crack driving force (the form of $\partial \psi(\phi,\mathbf{\varepsilon})/\partial \phi$ employed in the phase field evolution equation) and contact model (the form of $\partial \psi(\phi,\mathbf{\varepsilon})/\partial \mathbf{\varepsilon}$ employed in the mechanical equilibrium equation) is the key to arriving at an acceptable mechanics formulation based on the tension-compression splits of $\psi(\mathbf{\varepsilon})$ that are commonly used in the literature.
The simulations of in-plane shear (mode II) fracture of a homogeneous domain demonstrate that the crack driving force corresponding to the strain-spectral split of $\psi(\mathbf{\varepsilon})$ is in reasonable agreement with canonical experimental results and therefore the sharp-crack theories of fracture that are based upon them \cite{erdogan_crack_1963,zehnder_fracture_2012}.
The other crack driving forces we consider, the isotropic model with no splitting and the volumetric-devatoric split, do not pass this simple test.

The agreement between the experimental crack path and our mode II fracture simulations with the strain-spectral crack driving force holds for all cases except one, Fig.\ \ref{fig:shear_pf_compare}a, with the variational strain-spectral model and a phase field initial crack.
This exception is the only case out of Figs.~\ref{fig:shear_pf_compare} and \ref{fig:shear_cx_compare} in which the strain-spectral contact model is active and exposed to non-tensile strains.
The fixed contact resulting from this model significantly alters the stress distribution for mode II loading compared to the stress-free contact model, as shown in Fig.\ \ref{fig:shear_stress}, which likely results in the disagreement seen in Fig.\ \ref{fig:shear_pf_compare}a.
The strain-spectral contact model only weakens the material completely to purely tensile strains, resulting in significant stresses for shear strains.
Cracks that are not orthogonal to the tensile loading direction (i.e., horizontal in Fig.\ \ref{fig:paths_mixed_elasticity}) are likely to contain shear strains, leading to the higher stresses for the strain-spectral contact model noted in Fig.~\ref{fig:stress_strain_mechanics}b.
We anticipate that these exaggerated `frictional' stresses lead to the artificial widening and smoothing of cracks in heterogeneous structures observed in Figs.~\ref{fig:paths_mixed_elasticity} and \ref{fig:paths_alternating_grf}.

The two models in which both the crack driving force and contact model are satisfactory are both non-variational: they combine the strain-spectral crack driving force with either the stress-free contact model or the volumetric-deviatoric contact model.
One would expect to find differences between these two models under compression, where the contact model is likely to have a greater effect on the stress distribution and overall stiffness of the structure.
For the load cases we consider here, however, we observe no significant differences in crack paths or stress-strain curves between the non-variational models.
The stress-free contact model is computationally advantageous because it is linear \cite{ambati_review_2015}, and thus we use it with the strain-spectral crack driving force for the simulations in other sections of this paper.

It is perhaps unsatisfying that our results favor non-variational models.
While such models are increasingly popular \cite{ambati_review_2015,bilgen_crack-driving_2019} and can more easily match empirical strength surfaces for macroscopically homogeneous materials \cite{wu_unified_2017,kumar_revisiting_2020}, variational models have an appealing theoretical coherence.
We suspect that the free energy functionals that have been proposed for phase field fracture are intrinsically too simple to correctly model contact in a variational model because they lack information about the orientation of the crack.
The crack/surface normal vector is essential to models of static friction such as Coloumb's law \cite{popov_coulombs_2017}, which we presume to be the desired physics for a non-healing crack after fracture.
In contrast, phase field formulations determine stress-strain response based purely on the state of strain and the pointwise value of the phase field.
Such formulations lack the angular information provided by the crack normal vector, and thus will only ever coincidentally match models for frictional or frictionless contact.
In the absence of a unified variational model that captures both contact and fracture, it becomes a reasonable strategy to mix and match the parts of existing models that are least objectionable for the task at hand.

Finally, we consider our results in the context of other efforts to critically examine mechanics formulations.
Works that focus on strength surfaces \cite{kumar_revisiting_2020,de_lorenzis_nucleation_2021} are largely orthogonal to ours.
However, comparisons of crack paths from mode II fracture simulations are provided in Refs.~\cite{ambati_review_2015,bilgen_crack-driving_2019,zhang_assessment_2022}.
Curiously, these studies find different crack paths for the strain-spectral model: Refs.~\cite{ambati_review_2015,zhang_assessment_2022} found a crack path similar to our stress-free initial crack, while Ref.~\cite{bilgen_crack-driving_2019} found one similar to our phase field initial crack (i.e., in poor agreement with experiments).
For Ref.~\cite{zhang_assessment_2022}, it appears that a stress-free initial crack was in fact used, but the choice of initial crack is not given explicitly in Refs.~\cite{ambati_review_2015,bilgen_crack-driving_2019}.
In additional simulations, we found moderate effects on the mode II crack path from the length of the initial crack and the Poisson's ratio that do not affect our conclusions, but which would affect qualitative comparisons to other works.
The use of periodic vs.~fixed boundary conditions could also have an effect, but this is difficult to check with our methods.
Due to the lack of heterogeneity, we do not expect effects from other differences in formulation (e.g., AT1 vs.~AT2 and near-equilibrium vs.~minimization).

\subsection{Phase Field Formulation}

In this section, we consider effects of three aspects of phase field fracture models that have no direct equivalent in models for propagation of sharp cracks: the form of the pointwise fracture energy density $f(\phi)$ (AT1 vs.\ AT2), the choice of irreversibility criterion for $\phi$ (crack-set vs.\ damage), and the ratio between the microstructural length scale $L_\mathrm{cut}$ and the crack width parameter $\ell$.
Effects of these aspects of the model have been extensively considered for homogeneous materials (see, e.g., Refs.~\cite{pham_gradient_2011,tanne_crack_2018} for AT1 vs.\ AT2 and Ref.~\cite{tanne_crack_2018} for effects of the irreversibility condition and $\ell$).
Our focus is therefore on qualitative differences in crack paths in randomly heterogeneous structures.
Simulations in this section are conducted with the near-equilibrium evolution method using a uniaxial tensile applied strain with the increment given in Eq.~\eqref{eq:loading_tensile}.
We focus primarily on the strain-spectral/stress-free mechanics model, but cross-effects with other mechanics models are also illustrated.

\subsubsection{Model Comparison}
\begin{figure}
	\centering
	\includegraphics[width = 16.0cm]{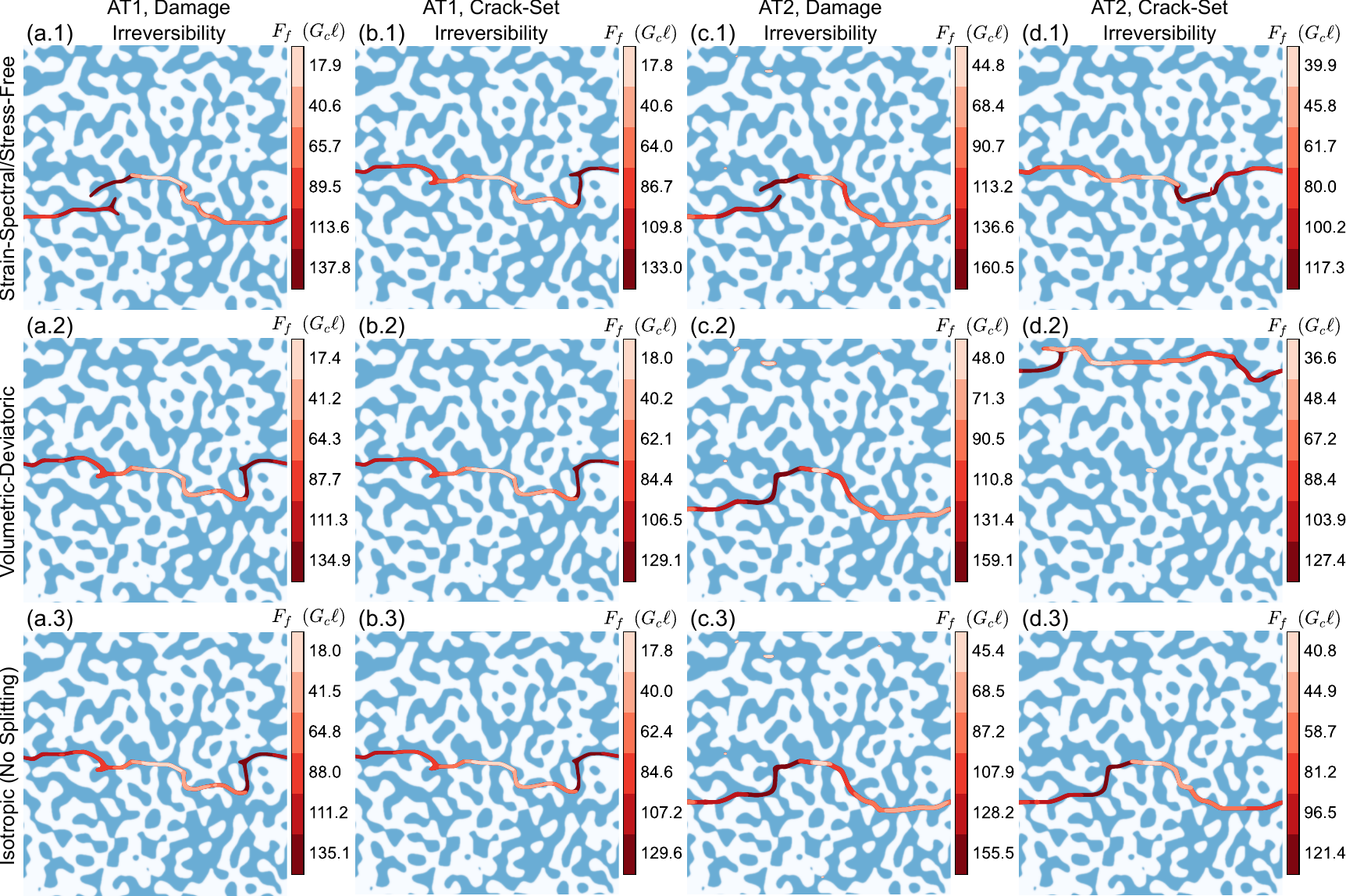}
	\caption{Crack paths for the two-phase structure from Fig.~\ref{fig:paths_mixed_elasticity} simulated under a uniaxial tensile applied strain with the (a,b) AT1  and (c,d) AT2 phase field formulations with (a,c) damage and (b,d) crack-set  irreversibility criteria using the (1) strain-spectral/stress-free, (2) volumetric-deviatoric, and (3) isotropic mechanics models.
}
	\label{fig:paths_formulation}
\end{figure}

Figure \ref{fig:paths_formulation} considers how crack paths are affected by three different aspects of the model formulation: AT1 vs.\ AT2, damage vs.\ crack-set irreversibility, and the choice between the isotropic, volumetric-deviatoric, and strain-spectral/stress-free mechanics models.
(We exclude the strain-spectral variational model due to its non-physical crack widening and the strain-spectral/vol.-dev.\ model due to its similarity to the strain-spectral/stress-free model.)
Figure \ref{fig:paths_formulation} demonstrates a striking sensitivity of the crack path to all three aspects of the formulation.
For the strain-spectral/stress-free model in Fig.~\ref{fig:paths_formulation}a-d.1, we see three different crack paths, with only the AT1 damage and AT2 damage cracks closely resembling each other (Fig.~\ref{fig:paths_formulation}a.1 and d.1, respectively).
However, even these two crack paths have different intermediate states and different final values for the fracture energy $F_f$, with $137.8G_c\ell$ and $160.5G_c\ell$ for the AT1 and AT2 damage cases, respectively.
Considering the other mechanics models, we see at least five distinct crack paths, with Fig.~\ref{fig:paths_formulation}a.1, b.1, d.1, c.2, and d.2 being typical examples.
Combined with results for other structures (not included here), Fig.\ \ref{fig:paths_formulation} suggests that there is no clear pattern for how these three aspects of the model formulation affect the crack path.
Between similar crack paths, we note that the damage-type irreversibility results in higher final fracture energies $F_f$ than the crack-set irreversibility (within Fig.~\ \ref{fig:paths_formulation}, compare a.2 to b.2, or c.3 to d.3, for example), and the AT2 damage model in particular usually has the highest values of $F_f$ overall.

\begin{figure}
	\centering
	\includegraphics[width = 9cm]{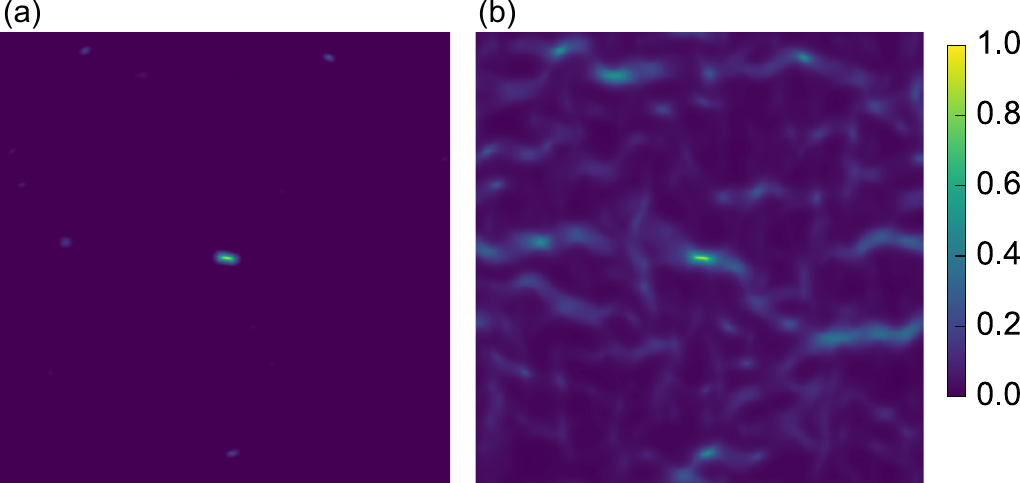}
	\caption{Pseudocolor images of the phase field $\phi$ at simulation steps corresponding to fracture initiation with (a) the AT1 model and (b) the AT2 model for the two-phase structure depicted in Fig.~\ref{fig:paths_formulation}.
}
	\label{fig:phi_formulation}
\end{figure}

To illustrate a key difference between the AT1 and AT2 models, Fig.~\ref{fig:phi_formulation} shows the phase field during fracture initiation in simulations corresponding to Fig.~\ \ref{fig:paths_formulation}a.1 and d.1.
Fig.~\ref{fig:phi_formulation}a and b thus correspond to the AT1 damage and AT2 crack-set models, respectively, but the irreversibility condition should not affect the distribution of $\phi$ prior to fracture initiation.
In the AT1 model, evolution of $\phi$ is localized to a peak at the primary nucleation site in the center and 4-5 smaller peaks in other locations, while the rest of the structure is undamaged.
In the AT2 model, $\phi$ is non-zero within the entire structure, and broad regions exist with moderate damage ($0.2 < \phi < 0.6$).
With crack-set irreversibility, these regions have the opportunity to `heal' after fracture initiation, but with damage irreversibility they affect the structure permanently.
One of these regions in the lower right of Fig.~\ref{fig:phi_formulation}b matches the location of a secondary crack in the AT2 damage case in Fig.~\ref{fig:paths_formulation}c.1 that eventually merges with the primary crack.

\begin{figure}
	\centering
	\includegraphics[width = 16cm]{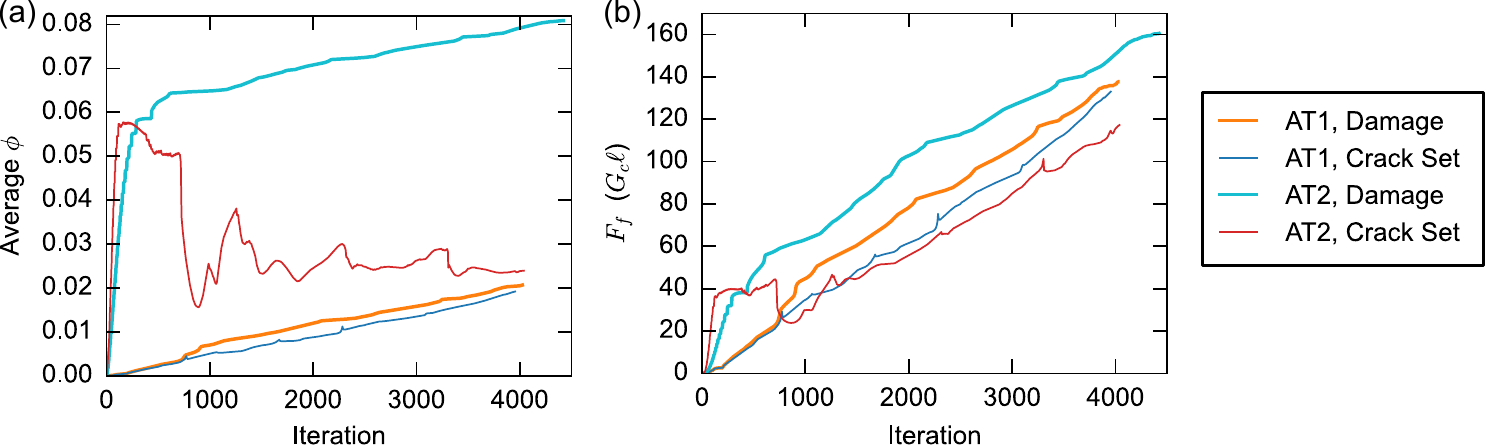}
	\caption{Plots of (a) average $\phi$ and (b) fracture energy $F_f$ vs.\ iteration for simulations of fracture of a two-phase structure with the AT1 model with crack-set (thin blue line) and damage (thick orange line) irreversibility and the AT2 model with crack-set (thin red line) and damage (thick cyan line) irreversibility.  Simulations plotted correspond to the crack paths in Fig.~\ref{fig:paths_formulation}a-d.1.
}
	\label{fig:iteration_formulation}
\end{figure}

To understand differences in evolution over the course of the entire simulation, Fig.~\ref{fig:iteration_formulation} plots evolution of the average value of $\phi$ and the  fracture energy $F_f$ vs.\ iteration for the simulations of two-phase structures shown in Fig.~\ref{fig:paths_formulation}a-d.1.
In Fig.~\ref{fig:iteration_formulation}a, simulations with the AT2 model show a large initial  increase in average $\phi$, which should correspond to the state shown in Fig.~\ref{fig:phi_formulation}b.
With the crack-set irreversibility condition, this increase is followed by a large decrease and additional oscillations until the end of the simulation.
With the damage irreversibility condition, average $\phi$ continues to increase monotonically at a slower rate, leading to a very high final average $\phi$ compared to the other three cases.
With the AT1 model, both irreversibility conditions result in similar steady growth, with occasional slight decreases in average $\phi$ observed for crack-set irreversibility.
Compared to average $\phi$, there is less of a difference between AT1 and AT2 in the evolution of $F_f$ in Fig.~\ref{fig:iteration_formulation}b.
This is consistent with the delocalized evolution of $\phi$ in Fig.~\ref{fig:phi_formulation}b because low values of $\phi$ contribute less to $F_f$ in the AT2 model due to the quadratic form of $f(\phi)$.
All four cases show steady increases in $F_f$, with monotonic growth for damage irreversibility and occasional slight decreases for crack-set irreversibility.

\begin{figure}
	\centering
	\includegraphics[width = 13cm]{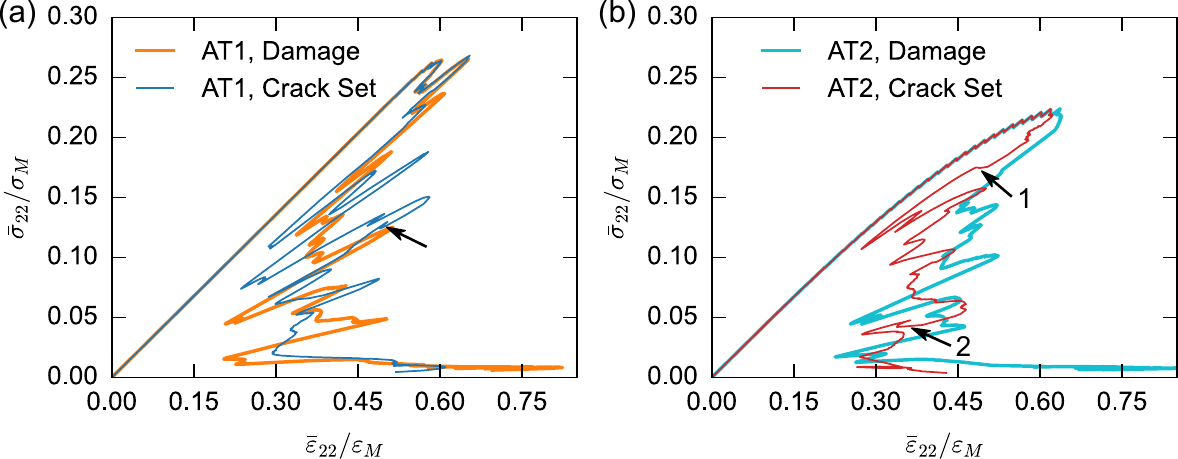}
	\caption{Stress-strain plots for fracture of a two-phase structure with (a) the AT1 model with crack-set (thin blue line) and damage (thick orange line) irreversibility and (b) the AT2 model with crack-set (thin red line) and damage (thick cyan line) irreversibility.  Simulations plotted correspond to the crack paths in Fig.~\ref{fig:paths_formulation}a-d.1.  Selected instances of `stiffening' (increases in average stiffness) with crack-set irreversibility are highlighted with black arrows.  The scaling stress $\sigma_M$ for both plots is based on the AT1 model, i.e., Eq.~\eqref{eq:sigma_M}.
}
	\label{fig:stress_strain_formulation}
\end{figure}

Evolution of $\phi$ in the AT2 model prior to fracture is well-known to affect mechanical response \cite{pham_gradient_2011}.
Figure \ref{fig:stress_strain_formulation} shows stress-strain plots for the simulations in Fig.~\ref{fig:paths_formulation}a-d.1.
In Fig.~\ref{fig:stress_strain_formulation}b, the AT2 model results in a decrease in stiffness prior to fracture.
This in turn results in a lower fracture stress of $0.22\sigma_M$ compared to $0.27\sigma_M$ for the AT1 model in Fig.~ \ref{fig:stress_strain_formulation}a, with $\sigma_M$ calculated for the AT1 model from Eq.~\eqref{eq:sigma_M}.
Sawtooth-like features prior to fracture with the AT2 model corresponds to relaxation of $\phi$ before the next strain increment is applied.
Stress-strain curves for the two irreversibility conditions are the same prior to fracture, but eventually they deviate due in part to the differences in crack path shown in Fig.~\ref{fig:paths_formulation}.
Stress-strain curves for the crack-set cases show signs of `stiffening' (increases in average stiffness), likely due to healing of $\phi$ where it is below the crack-set threshold of 0.9.
At the location indicated `1' in Fig.~\ref{fig:stress_strain_formulation}b, this stiffening occurs during a decrease in applied strain, resulting in a nearly horizontal segment of stress-strain curve.
At location 2, stiffening coincides with an increase in applied strain, resulting in a snap-back event with a cusp appearing `inside' another snap-back event.
Our control algorithm for the near-equilibrium method handles these examples gracefully, but in general additional precautions may be needed to prevent simulations with crack-set irreversibility from being trapped in cycles of loading and unloading that lack irreversible evolution.

\subsubsection{Convergence with respect to microstructural length scale}
We now consider how differences between the AT1 and AT2 models and the damage and crack-set irreversibility conditions change as the microstructural length scale $L_\mathrm{cut}$ is increased relative to the crack width parameter $\ell$.
This can be interpreted as an evaluation of $\Gamma$-convergence, since our use of $\ell$ as a characteristic length scale prevents us from investigating the limit $\ell\to 0$ directly.
As noted in Section \ref{sec:structure_gen}, we change $L_\mathrm{cut}$ by interpolating structures generated with $L_\mathrm{cut} =6\ell$ in a domain of size $L_x=L_y=100\ell$ onto a larger grid with $1023^2$ or $2047^2$ points compared to the original grid size of $N_x=N_y=511$.
These larger grids in turn represent larger domain sizes, $L_x=L_y=200\ell$ or $L_x=L_y=400\ell$, resulting in $L_\mathrm{cut}=12\ell$ or $L_\mathrm{cut}=24\ell$, respectively.

\begin{figure}
	\centering
	\includegraphics[width = 16.4cm]{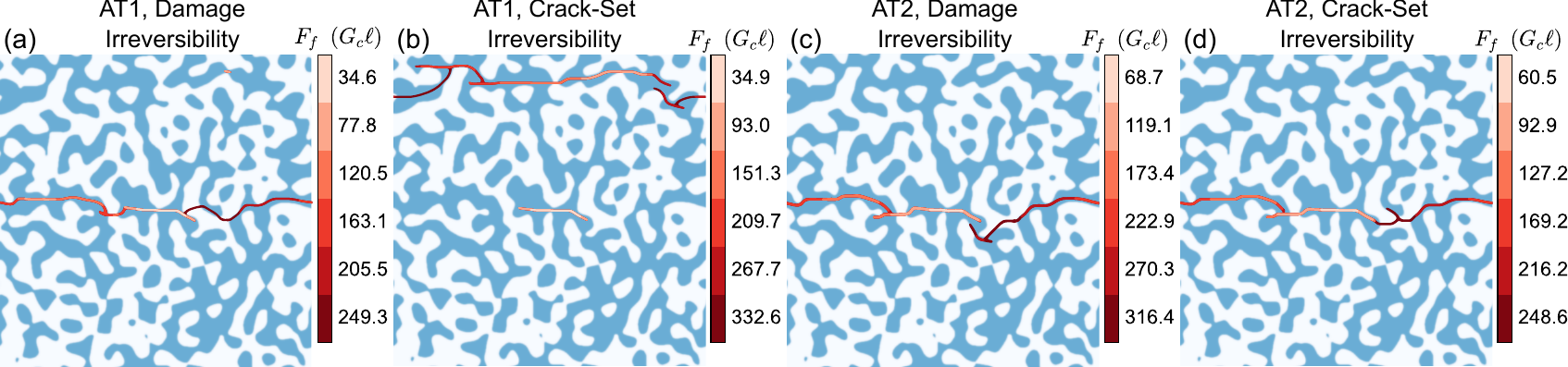}
	\caption{Crack paths for the structure from Fig.~\ref{fig:paths_formulation} upscaled to size $L_x=L_y=200\ell$ such that the cutoff length scale $L_\mathrm{cut}$ describing the microstructure is $12\ell$ instead of $6\ell$. Simulation conditions are otherwise the same as Fig.~\ref{fig:paths_formulation}a-d.1.
}
	\label{fig:paths_ell}
\end{figure}

Figure \ref{fig:paths_ell} shows crack paths resulting from simulations under the same conditions as in Fig.~\ref{fig:paths_formulation}a-d.1 with the structure upscaled from $L_\mathrm{cut}=6\ell$ to $L_\mathrm{cut}=12\ell$.
These crack paths are thinner than their equivalents with $L_\mathrm{cut}=6\ell$ (a result of our use of the $\phi=0.5$ contour for visualization), and exhibit sharp changes in direction that might have appeared smoother in similar cracks at $L_\mathrm{cut}=6\ell$.
In their overall structure, three of the crack paths (corresponding to the AT2 model and the AT1 damage case) now agree with each other for much of their length.
The evolution of these crack paths most closely resembles the AT2 crack-set case in Fig.~\ref{fig:paths_formulation}d.1, while the final crack path also resembles the AT1 crack-set case in Fig.~\ref{fig:paths_formulation}b.1 and similar crack paths with the AT1 model and other mechanics formulations.
Meanwhile, the AT1 crack-set case in Fig.~\ref{fig:paths_ell}b is changed significantly from Fig.~\ref{fig:paths_formulation}b.1: the initial crack now stops growing and the structure is fractured by a secondary crack initiated in the upper right corner.
Consistent with Fig.~\ref{fig:paths_ell}, results for other realizations of the two-phase structure show more agreement between crack paths with different model formulations at $L_\mathrm{cut}/\ell=12$ than at $L_\mathrm{cut}/\ell=6$.
Compared to Fig.~\ref{fig:paths_ell}, these realizations result in agreement between $L_\mathrm{cut}/\ell=12$ and $L_\mathrm{cut}/\ell=6$ with the same model formulation more often than is indicated by Fig.~\ref{fig:paths_ell} alone.
However, there does not appear to be any pattern in this agreement between forms of $f(\phi)$ or irreversibility conditions.

\begin{figure}
	\centering
	\includegraphics[width = 13.6cm]{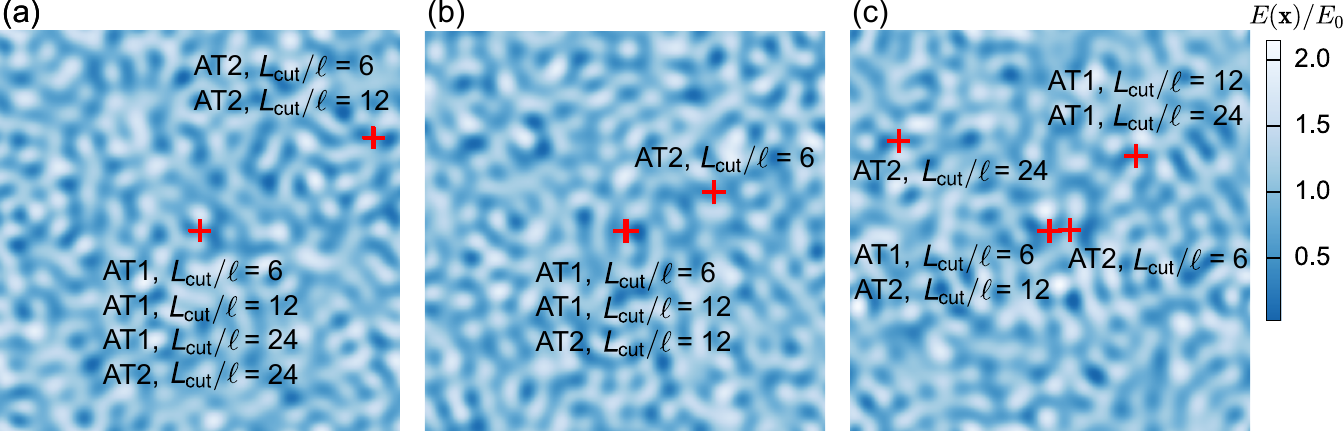}
	\caption{Comparison of crack nucleation sites, indicated by red plus signs, between the AT1 and AT2 models for three smooth random structures (a, b, and c) that are upscaled to create three size scales: $L_\mathrm{cut}=6\ell$, $L_x=L_y=100\ell$ (original); $L_\mathrm{cut}=12\ell$, $L_x=L_y=200\ell$; and $L_\mathrm{cut}=24\ell$, $L_x=L_y=400\ell$.  The nucleation location appears to converge at lower values of $L_\mathrm{cut}/\ell$ for the AT1 model compared to the AT2 model.
}
	\label{fig:nucleation_grf}
\end{figure}

A clearer picture emerges when $L_\mathrm{cut}/\ell$ is increased in the smooth random structures.
These do not exhibit much variation in crack path with the near-equilibrium evolution method (most are nearly flat, as in Fig.~\ref{fig:paths_small_evolution}c), but the location where the crack nucleates can differ between the AT1 and AT2 models.
Figure \ref{fig:nucleation_grf} compares nucleation sites between the AT1 and AT2 models for three smooth random structures at two or three levels of $L_\mathrm{cut}/\ell$.
Nucleation sites in Fig.~\ref{fig:nucleation_grf} are designated as the location where $\phi$ first exceeds 0.9 in a given simulation.
In all three of the original structures with $L_\mathrm{cut}=6\ell$, the AT2 model nucleates the crack at a different location from the AT1 model.
(For comparison, this was only observed in one of the three two-phase structures for $L_\mathrm{cut}=6\ell$.)
In two out of three cases, the nucleation site for the AT2 model converges to that of the AT1 model as $L_\mathrm{cut}/\ell$ increases, with agreement at  $L_\mathrm{cut}/\ell=24$ in Fig.~\ref{fig:nucleation_grf}a and $L_\mathrm{cut}/\ell=12$ in Fig.~\ref{fig:nucleation_grf}b.
In Fig.~\ref{fig:nucleation_grf}c, the nucleation site of the AT1 model is itself not converged at $L_\mathrm{cut}/\ell=6$, but the AT2 model still nucleates at this `old' site at $L_\mathrm{cut}/\ell=12$, before nucleating at yet another site when $L_\mathrm{cut}/\ell=24$.
 The AT1 model in Fig.~\ref{fig:nucleation_grf} uses damage irreversibility while the AT2 model uses crack-set irreversibility, but we do not expect the irreversibility condition to affect the nucleation site.

\subsubsection{Discussion}
One way to interpret the differences between crack paths in Fig.~\ref{fig:paths_formulation} is that one model (and thus crack path) is more correct than the others.
For the mechanics formulations and evolution methods, we evaluated models based in part on simple simulations that are easier to analyze than the heterogeneous structures.
In this section, we primarily refer to analyses already present in the literature.

Linse et al.~\cite{linse_convergence_2017} find that damage irreversibility prevents convergence of the dissipated fracture energy $F_f$ to its ideal value because diffuse evolution of $\phi$ prior to nucleation of the crack is not allowed to heal.
This diffuse evolution of $\phi$ is not eliminated by decreasing $\ell$ relative to the length of a reduced-stiffness region in their 1D domain, and thus they find that the damage irreversibility condition is incompatible with $\Gamma$-convergence of $F_f$.
In our own results, we find a difference $F_f$ between the damage and crack-set irreversibility conditions for the AT2 model (e.g., in Fig.~\ref{fig:paths_formulation}) that matches the findings of Linse et al., but it is not clear whether the damage irreversibility condition harms $\Gamma$-convergence of the crack path (in our case, convergence in the limit $L_\mathrm{cut}/\ell \to \infty$).
In the AT1 model, which they do not consider, we find a much smaller effect of damage irreversibility on $F_f$ compared to the AT2 model.

Tann\'e et al.~\cite{tanne_crack_2018} fixed $\ell$ based on the fracture stress and the toughness $G_c$ and found that the AT1 model successfully approximates fracture across a range of weak and strong stress singularities (introduced by V-notches) and concentrations (introduced by U-shaped notches).
The AT2 model was found to successfully model crack nucleation at strong stress concentrations/singularities, but it diverged from experiments for weaker concentrations/singularities.
This agrees quite well with our experience, where the two-phase structures have stronger stress concentrations than the smooth random structures (per the lower fracture stress and strain in Fig.~\ref{fig:stress_strain_evolution}b compared to Fig.~\ref{fig:stress_strain_evolution}a) and result in less of a difference in nucleation behavior between the AT1 and AT2 models.
In the smooth random structures, faster convergence of the nucleation location with respect to $L_\mathrm{cut}/\ell$  for the AT1 model supports its use instead of the AT2 model for weak stress concentrations even when $\ell$ is not fixed based on material properties.
(The relationship between $\ell$, $G_c$, and fracture stress can be manipulated by changing $f(\phi)$ and $h(\phi)$, for example \cite{wu_length_2018}.)

Overall, our findings are generally consistent with those of Tann\'e et al.~\cite{tanne_crack_2018} and Linse et al.~\cite{linse_convergence_2017}, which taken together suggest that the AT1 model and crack-set irreversibility should be preferred.
However, the high sensitivity of crack paths in the two-phase structure to any change in model formulation suggests that focusing on a single `correct' path may not be a desirable approach.
For one thing, none of the methods is consistently converged with respect to $L_\mathrm{cut}/\ell$ for $L_\mathrm{cut} < 12\ell$, suggesting that studying only converged `correct' paths may be computationally challenging.
Furthermore, the crack paths shown in Figs.~\ref{fig:paths_formulation} and \ref{fig:paths_ell} are all qualitatively similar, with no clear systematic difference due to the irreversibility condition, form of $f(\phi)$, or mechanics formulation.
Even increasing $L_\mathrm{cut}/\ell$ only appears to systematically affect the crack path at small length scales close to the crack width.
If crack propagation in the two-phase structures is interpreted as a highly sensitive or chaotic process \cite{gerasimov_stochastic_2020}, then all of these aspects may simply be influencing which crack path is selected out of several statistically indistinguishable realizations.
The real question then becomes which aspects of model formulation have systematic effects on statistical descriptors of the crack path, such as its power spectrum \cite{ponson_statistical_2016}.
This question is substantially different from the qualitative approach taken in this work; we speculate on which aspects of phase field fracture models will result in such quantitative effects in the following section.

\subsection{Discussion}

In Section \ref{sec:background}, we examined several different ways in which phase field models for quasi-static brittle fracture can be formulated.
In the previous three sections, we have tested the effects of different formulations on crack path selection in elastically heterogeneous microstructures.
Our results indicate that the near-equilibrium evolution method and non-variational mechanics models with the strain-spectral driving force are better than their alternatives at modeling quasi-static brittle fracture.
We argue based on our simulation results and previous studies that the AT1 model should be preferred to the AT2 model and the crack-set irreversibility condition to the damage irreversibility condition.
Crack path selection in our two-phase structures appears to be highly sensitive to the aspects of model formulation that we consider, and an `ideal' model for quasi-static brittle fracture under tension would be composed of these preferred variants.

However, it is not clear whether many of the differences in crack path result from differences in the formulations themselves or from the high sensitivity of the crack path selection process in the two-phase structures.
Quantitative analysis of a larger dataset of crack paths for an ensemble of statistically identical microstructures would provide a stronger basis for evaluating systematic differences between model variants, and would be an interesting approach for future work.
This would be in a way similar to a stochastic approach proposed by Gerasimov et al.~\cite{gerasimov_stochastic_2020} for systems with homogeneous material properties.
In the meantime, we can speculate about quantitative effects based on our qualitative observations.
We expect that sufficiently high overstress would result in systematic and statistically significant effects on crack paths simulated with the time-dependent and minimization evolution methods in heterogeneous structures.
Likewise, the crack smoothing and widening observed for the strain-spectral variational mechanics formulation systematically affects the crack path and the dissipated fracture energy $F_f$.
Damage irreversibility in the AT2 model also systematically affects $F_f$, although we cannot yet say whether it systematically affects the crack path.
A minimal recommendation from this study is that these expected systematic effects on crack path and $F_f$ should be avoided when investigating randomly heterogeneous materials.

\section{Conclusions}
We have presented a comprehensive overview of how popular variants of the phase field model for quasi-static brittle fracture affect crack path selection in systems with both uniform and randomly heterogeneous elastic properties.
We consider four ways in which phase field models for quasi-static brittle fracture can vary: in how the phase field is evolved, in the formulation of the coupling between the elastic and phase field, the form of the phase field approximation to the crack length/area (AT1 or AT2 \cite{tanne_crack_2018}), and the conditions under which evolution of the phase field is considered to be irreversible (everywhere, as in damage models, or only within a crack set).
We probe the effects of these variants in simulations with spatially uniform elastic properties, random two-phase structures with contrasting Young's moduli, and random structures with smoothly varying Young's modulus.
For the random structures, we examine how crack paths (and their sensitivity to model variants) change as the size scale of the structure is changed relative to the crack width parameter $\ell$ in the phase field model.

We consider all of these variants within a common numerical approach that combines novel and standard aspects.
We identify three types of evolution method for the phase field: minimization, time evolution, and near-equilibrium evolution.
Our implementation of the near-equilibrium method is novel but the others are standard, and for all three methods we employ staggered solutions of the phase field and mechanics sub-problems.
We use an FFT-accelerated strain-based micromechanics solver for the mechanics sub-problem~\cite{zeman_finite_2017,leute_elimination_2022,ladecky_optimal_2022}, and implement a bound-constrained conjugate gradients algorithm~\cite{vollebregt_bound-constrained_2014} to solve for the phase field while directly enforcing irreversibility constraints.

We find that crack paths in heterogeneous structures differ significantly between the near-equilibrium evolution method and the minimization and time-dependent evolution methods under overstressed conditions.
Such conditions occur when the near-equilibrium method undergoes unloading (i.e., snap-back) but the time-dependent and minimization methods do not.
This effect on crack path in the minimization method relies on the interaction between overstress and material heterogeneity, and thus it is not apparent in simulations without heterogeneity.
Effects of overstress with the time-dependent method, such as crack broadening and branching, resemble effects of overstress in dynamic fracture models \cite{bleyer_dynamic_2017}.
The near-equilibrium method most closely resembles classical models for quasi-static fracture and avoids overstress-related effects present in the other methods.

In our examination of different mechanics formulations, we find distinct effects due to the choice of driving force (elasticity formulation in the phase field governing equation) and the choice of contact model (elasticity formulation in the mechanical equilibrium equation).
We consider elastic energy densities with no splitting between tension and compression (i.e., the isotropic model), and splittings based on volumetric-deviatoric and spectral decompositions of the strain tensor.
Of these, only the driving force for the strain-spectral split results in agreement with an experimental mode II (in-plane shear) crack path.
However, the contact model for the strain-spectral split results in a crack that bears significant shear stresses, leading to artificial widening and smoothing of cracks in heterogeneous microstructures.
Desirable combinations of driving force and contact model for predominately tensile loading can be obtained via non-variational mechanics formulations that combine the driving force from the strain-spectral split with the contact model from another formulation.

We find that crack paths in the heterogeneous structures are sensitive to multiple aspects of the formulation besides evolution method and mechanics formulation.
Sensitivity to aspects that do not have an equivalent in sharp-crack models of crack propagation (e.g., AT1 vs.~AT2 and crack-set vs.~damage irreversibility) is reduced when the length scale of the microstructure is larger relative to the crack width parameter $\ell$.
Our results along with previous work \cite{tanne_crack_2018} suggest that the AT1 model is advantageous for $\Gamma$-convergence in the presence of relatively weak stress concentrations.
Likewise, the crack-set irreversibility condition is preferable for use with the AT2 model \cite{linse_convergence_2017}.
A potential approach for future studies would be to test for systematic differences between methods via statistical characterization of the crack path.

\section*{Acknowledgements}

We thank Till Junge and Jan Zeman for useful discussion, and Ali Falsafi, Richard Leute, Antoine Sanner, and Sindhu Singh for assistance in code development and deployment. We used \textsc{$\mu$Spectre} (\url{https://gitlab.com/muspectre/muspectre}) to solve the mechanical problem. Funding was provided by the European Research Council (StG-757343) and by the Deutsche Forschungsgemeinschaft under Germany’s Excellence Strategy (EXC-2193/1 – 390951807). Numerical simulations were performed on bwForCluster NEMO (University of Freiburg, DFG grant INST 39/963-1 FUGG).

\section*{Competing interests}
The authors declare that they have no known competing financial interests or personal relationships that could have appeared to influence the work reported in this paper.

\bibliography{references}

\end{document}